\documentclass[aps,reprint]{revtex4-1}
\usepackage[usenames,dvipsnames]{color}
\usepackage{amsmath,array,amssymb}
\usepackage{graphicx}
\usepackage{hyperref}
\usepackage{textcomp}
\usepackage{braket}
\usepackage{multirow}
\usepackage{indentfirst}
\usepackage[FIGTOPCAP,raggedright,nooneline,bf,footnotesize]{subfigure}
\usepackage{paralist}
\usepackage{overpic}
\usepackage{soul} 
\usepackage{xspace} 

\usepackage{comment}
\usepackage{tablefootnote}
\usepackage{colortbl}
\definecolor{LightCyan}{rgb}{0.88,1,1}
\definecolor{Plum}{rgb}{.95,.7,.95}

\newcommand{\cmt}[1]{}

\renewcommand{\vec}[1]{\boldsymbol{#1}}
\newcommand{\figref}[1]{Fig.~\ref{#1}}

\renewcommand{\eqref}[1]{Eq.~(\ref{#1})}

\newcommand{\eryso}[0]{{Er$^{3+}$:Y$_2$SiO$_5$}\xspace}
\newcommand{\ndyso}[0]{{Nd$^{3+}$:Y$_2$SiO$_5$}\xspace}
\newcommand{\yso}[0]{Y$_2$SiO$_5$\xspace}
\newcommand{\er}{Er$^{3+}$\xspace}
\newcommand{\nd}{Nd$^{3+}$\xspace}
\newcommand{\y}{Y$^{3+}$\xspace}

\newcommand{\cawo}[0]{CaWO$_4$\xspace}
\newcommand{\linbo}[0]{LiNbO$_3$\xspace}
\newcommand{\ercawo}[0]{{Er$^{3+}$:CaWO$_4$}\xspace}
\newcommand{\erlinbo}{Er$^{3+}$:LiNbO$_3$\xspace}

\graphicspath{{img/}}

\begin{document}

\title{Optical study of the anisotropic erbium spin flip-flop dynamics}

\author{B. Car$^1$}
\author{L. Veissier$^1$}
\author{A. Louchet-Chauvet$^1$}
\author{J.-L. Le Gou\"et$^1$}
\author{T. Chaneli\`ere$^{1,2}$}
\affiliation{$^1$Laboratoire Aim\'e Cotton, CNRS, Univ. Paris-Sud, ENS-Cachan, Universit\'e Paris-Saclay, 91405, Orsay, France}
\affiliation{$^2$Univ. Grenoble Alpes, CNRS, Grenoble INP, Institut N\'eel, 38000 Grenoble, France}

\date{\today}

\begin{abstract}
We investigate the erbium flip-flop dynamics as a limiting factor of the electron spin lifetime and more generally as an indirect source of decoherence in rare-earth doped insulators.
Despite the random isotropic arrangement of dopants in the host crystal, the dipolar interaction strongly depends on the magnetic field orientation following the strong anisotropy of the $g$-factor. In \eryso, we observe by transient optical spectroscopy a three orders of magnitude variation of the erbium flip-flop rate (10ppm dopant concentration). The measurements in two different samples, with 10ppm and 50ppm concentrations, are well-supported by our analytic modeling of the dipolar coupling between identical spins with an anisotropic $g$-tensor. The model can be applied to other rare-earth doped materials. We extrapolate the calculation to \ercawo, \erlinbo and \ndyso at different concentrations.

\end{abstract}


\maketitle

\section{Introduction}
Rare-earth ions are appealing qubits to store and process quantum information. When embedded in crystalline matrices, they combine well protected spin and optically active transitions. The interplay between the spin and atomic like optical excitations places the rare-earth species at the interface between solid-state and atomic physics. Within the lanthanide series, the 
Kramers ions (odd number of electrons) exhibit a strong paramagnetic sensitivity because of the crystal field levels admixture leading to a large anisotropic effective electron spin \cite{abragam2012electron}. They also hold a lot of promises in quantum information because a moderate bias magnetic field induces electron paramagnetic resonances (EPR) in the few GHz range. This typical splitting falls in the range of superconducting qubits to form a hybrid interface between the actively investigated quantum electronics circuits and solid-state spins as a memory buffer \cite{grezes, probst}. Additionally, the noticeably narrow inhomogeneous line ($\sim$ 1GHz) is well resolved at low field, thus allowing the spectral optical addressing of spin-selective transitions. The convergence of the appealing spin and optical properties has recently attracted a lot of attention to promote Kramers paramagnetic ions as a quantum spin-photon interface with highly diluted samples \cite{PhysRevLett.113.203601, o2014interfacing, fernandez2015coherent} or more recently with single ions \cite{dibos_isolating_2017, zhong2018optically}. Among them, erbium is emblematic because its optical transition falls in the telecom range and it exhibits large $g$-factor values ($g\sim 15$) in some 
crystals such as \eryso or \erlinbo \cite{bottger2006spectroscopy, sun2008magnetic,milori_optical_1995}. The tribute to pay to these attractive properties is a rapid loss of decoherence due to the enhanced sensitivity to the magnetic fluctuations. Whatever the viewing angle, optics or EPR, the spin dynamical properties are absolutely critical. They obviously directly impact the outcome of the EPR spectroscopy but they also influence the optical coherence time by the differential magnetic fluctuations between the ground and optically excited states \cite{bottger_decoherence}.

The spin dynamics is essentially governed by two mechanisms \cite{bottger_decoherence}. On the one side, spin lattice relaxations (SLR) describe the interaction of the spin with the phonon bath \cite{Orbach458}. They can be detailed in different mechanisms, namely direct, Raman and Orbach, depending on the process order (one or two-phonon) and the level diagram (on or off-resonance phonon interaction).
However, at liquid helium temperatures, the two-phonon processes are usually negligible. Thus, the SLR is dominated by the direct process. On the other side, the cross-relaxation between adjacent impurities with opposite spin orientations, also known as the flip-flop mechanism (FF), appears as an important process even in low doping samples \cite{van1948dipolar, portis1956spectral}. Both SLR an FF mechanisms primarily describing the population decay, or in other words the longitudinal relaxation, indirectly induce decoherence. Indeed, the fluctuations of the spins generate a background noise leading to decoherence, or transverse relaxation. SLR and FF have very different dependencies as a function of the applied magnetic field \cite{bottger_decoherence, saglamyurek2015efficient, Cruzeiro}. 
Practically, SLR dominates at large field and FF at low field. As the magnetic field energy splitting increases and becomes larger than $k_B T$ (where $k_B$ is the Boltzmann constant and $T$ the temperature), the spins become fully polarized, thus preventing the FF. At the same time, the phonon density increases, making the SLR more efficient. At the end, SLR and FF are well separated and can be considered and fought independently to prevent relaxation. 
At large magnetic fields, SLR can be controlled by adjusting the level structure with respect to the photon density. This can be done for example by a clever orientation of magnetic field relying on the highly anisotropic Zeeman interaction \cite{bottger2009effects, ranvcic2018coherence} or by a direct nanostructuration of the material to control the intrinsic phonon density\cite{Lutz, yang1999one, yang1999electron}. FF can be also avoided by cooling the spins to few tens of mK thus benefiting from a fully polarized sample at thermal equilibrium. By going deep into the SLR regime, the phonon-bottleneck relaxation has been observed very recently at 20 mK \cite{Budoyo}.

The low field limit at larger temperature, typically 2-4~K range, also deserves consideration because it offers relaxed experimental conditions. As previously mentioned, with a few GHz Zeeman splitting and well resolved optical transitions, this is a region of interest for quantum information applications. The FF dominates the spin dynamics in that case. A partial optical spin polarization can be obtained by frequency selective optical pumping, also known as spectral holeburning \cite{Hastings, saglamyurek2015efficient, Cruzeiro}. The imprinted spectral pattern, usually burned during the preparation stage of quantum memory protocols \cite[and references therein]{heshami2016quantum}, has a lifetime limited by the FF mechanism. Motivated by the potential applications and more generally by the possibility to optically control the out-of-equilibrium ensemble polarization, we optically study the FF cross-relaxation process between erbium spins.


The paper is organized as follows. We first remind the theoretical background to evaluate the FF rate between two identical spins. Even in the case of anisotropic $g$-tensors, we derive an analytical formula for the cross-relaxation lifetime. We then use optical techniques (transient excitation and accumulated spectral hole burning) to measure the anisotropy of the FF rate of \er in \yso as a function of the magnetic field orientation. We verify that the lifetime varies by several orders of magnitude as expected from the $g$-tensor theoretical calculation. We finally extend our analysis to another Kramers dopant \nd and to higher symmetry crystals (as \cawo and \linbo), which are also investigated for quantum information storage and processing.

\section{Theory}
\label{sec:theory}
The identical spins flip-flop dynamics has been investigated in the context of spin diffusion in a broad sense (spatial and spectral) \cite[chap.4, Spin diffusion in solids]{asakura1998solid}. Starting from the seminal work of Bloembergen \cite{BLOEMBERGEN1949386}, this is still an active subject of research widely stimulated by the perspectives in quantum information for which spin impurities in solids appear as a useful resource \cite[and references]{PhysRevApplied.6.044001}. We will follow the historical perturbative framework to describe the exponential relaxation of the out-of-equilibrium \er spin polarization.

\subsection{Flip-flop return to equilibrium}
\label{sec:theory_general}
We assume the $N$ spin-$1/2$ identical rare-earth impurity ions randomly substituted in host crystalline matrix. The energy level degeneracy is lifted by an external magnetic field $\vec{B}$. The resulting Zeeman splitting is set much smaller than the thermal energy $k_BT$ and, although weakly coupled to the heat bath, the spins are evenly distributed over the $\ket{+}$ and $\ket{-}$ eigenstates of the Zeeman Hamiltonian. The spins are coupled to each other by magnetic dipole-dipole interaction. The interaction strength is assumed to be much smaller than the Zeeman splitting, which only allows flip-flop transitions, where two spins simultaneously flip in opposite ways, with no total energy change.    
   
In a spectral hole burning experiment, a subset $W_+$, containing $N_0<<N$ impurity ions, can be prepared in state $\ket{+}$ at time $t=0$. That population imbalance is destroyed by FF interactions between the optically probed ions and the rest of the ensemble. The corresponding exponential-decay rate $R$ is given by the Fermi's Golden Rule \cite[W.B. Mims, Chap.4, p.294]{geschwind1972electron}:  
\begin{equation}
R =  \frac{2 \pi }{ \hbar}\left\{\sum_{j\in\left\{\ket{-}\right\}}\left|\bra{+-}\bf{H}_{ij}\ket{-+}\right|^2\right\}_{i\in W_+} \frac{1}{\hbar\Gamma^{\rm spin}_{\rm inh}} ,
\label{eq:FGR_rate}
\end{equation} 
where $\Gamma^{\rm spin}_{\rm inh}$ represents the inhomogeneous broadening of the spin transition, in $s^{-1}$, $\left\{\right\}_{i\in W_+}$ expresses averaging over the subset $W_+$, and $\bf{H}_{ij}$ stands for the magnetic dipole-dipole Hamiltonian of the $i,j$ ion pair:        
\begin{equation}
\bf{H}_{ij} = - \frac{\mu_0}{4\pi} \frac{1}{r_{ij}^3}  \left[ 3 \left( \vec{\mu}_{i} \cdot \vec{u}_{ij} \right) \left( \vec{\mu}_{j} \cdot \vec{u}_{ij} \right)-\vec{\mu}_{i} \cdot \vec{\mu}_{j} \right]  \;,
\label{eq:Hint}
\end{equation} 
where $\vec{r_{ij}}$, $\vec{u}_{ij}$, and $\vec{\mu}_{i,j}$ respectively represent the inter-ion vector, the unit vector along $\vec{r_{ij}}$, and the magnetic dipole moments. 

The $\bf{H}_{ij}$ structure suggests that the sum over $j$ in Eq.~\ref{eq:FGR_rate} can be approximately evaluated as a combination of angular averaging over $\vec{u}_{ij}$ direction and sum over the inter-ion distance. Therefore:
\begin{multline}
\sum_{j\in\left\{\ket{-}\right\}}\left|\bra{+-}\bf{H}_{ij}\ket{-+}\right|^2\approx\\
\left(\frac{\mu_0}{4\pi}\mu_{\rm B}^2\right)^2\times\Xi\left(\bar{\bar{g}},\vec{B}\right)\times\sum_{j\in\left\{\ket{-}\right\}} \frac{1}{r_{ij}^6}, 
\label{eq:sum_over_ions}
\end{multline} 
where $\mu_{\rm B}$ denotes the Bohr magneton, and the dimension-less coupling factor $\Xi\left(\bar{\bar{g}},\vec{B}\right)$, defined as:
\begin{multline}
\Xi\left(\bar{\bar{g}},\vec{B}\right)=\frac{1}{4\pi}\times\mu_{\rm B}^{-4}\times\\
\int d\vec{u}_{ij}\left|\bra{+-} 3 \left(\vec{\mu}_{i}\cdot\vec{u}_{ij}\right) \left(\vec{\mu}_{j} \cdot \vec{u}_{ij} \right)-\vec{\mu}_{i} \cdot \vec{\mu}_{j} \ket{-+}\right|^2,
\label{eq:Xi_factor}
\end{multline}
does not refer to a specific interacting ion pair but depends on the $\bar{\bar{g}}$-tensor, and on $\vec{B}$ direction, which determines the eigenvectors $\left(\ket{+},\ket{-}\right)$. The detailed calculation of $\Xi\left(\bar{\bar{g}},\vec{B}\right)$ is deferred to Appendix~\ref{sec:Vanalytical}. 
 
To evaluate the sum over $1/r_{ij}^6$ in Eq.~\ref{eq:sum_over_ions}, we replace the impurity-ion random distribution by an evenly spaced arrangement of $j$ spins in state $\ket{-}$ on a fictitious cubic lattice. Each $i$ spin in state $\ket{+}$ is thus surrounded by layers of those $j$ spins (see Appendix \ref{sec:sum_distance}).  Since the weighting factor $1/r_{ij}^6 $ rapidly drops with distance, one may truncate the sum to a few layers. With truncation to the first 10 layers, the sum reduces to $8.4n_s^2$, where $n_s$ represents the spatial density of spins in state $\ket{-}$, which coincides with half the impurity-ion density. Finally, the FF decay rate reads as:
\begin{equation}
R = \frac{2 \pi }{ \hbar}\left(\frac{\mu_0}{4\pi}\:n_s\:\mu_{\rm B}^2\right)^2\times 8.4 \times\Xi\left(\bar{\bar{g}},\vec{B}\right)\times\frac{1}{\hbar\Gamma^{\rm spin}_{\rm inh}} ,
\label{eq:FGR_rate_final}
\end{equation}      

A magnetic dipole moment $\vec{\mu}_i$ can be expressed in terms of the effective spin operator $\vec{\hat{S}}_i$ and of the $\bar{\bar{g}}$-tensor as:
\begin{equation}
\vec{\mu}_i=\mu_{\rm B}\bar{\bar{g}}\cdot\vec{\hat{S}}_i
\label{eq:mag_dipole}
\end{equation}
In the frame $(x,y,z)$ where the $\bar{\bar{g}}$-tensor is diagonal, $\vec{\mu}_i$ reads as:
\begin{equation}
\vec{\mu}_i=\mu_{\rm B}\left(\hat{S}_{i,x}g_x\vec{u}_x+\hat{S}_{i,y}g_y\vec{u}_y+\hat{S}_{i,z}g_z\vec{u}_z\right) 
\label{eq:mag_dipole_expansion}
\end{equation}
where:
\begin{equation}
\bar{\bar{g}} = \left( \begin{matrix} g_x & 0 & 0 \\ 0 & g_y & 0 \\ 0 & 0 & g_z \end{matrix} \right) \;
\end{equation}
and $(\vec{u}_x,\vec{u}_y,\vec{u}_z)$ represent the unit vectors of the reference frame. 
When $\vec{B}$ is directed along $Oz$, $\hat{S}_z$ and the Zeeman Hamiltonian share the same  eigenvectors. Hence, $\bra{+}\hat{S}_{i,z}\ket{-}$ vanishes and $\Xi\left(\bar{\bar{g}},\vec{B}\right)$ does not depend on $g_z$. Conversely, $\Xi\left(\bar{\bar{g}},\vec{B}\right)$ does not depend on $g_x$ or $g_y$ when $\vec{B}$ is directed along $Ox$ or $Oy$ respectively. That feature will prove to be important in the following. 

\subsection{Application to \eryso}
\label{sec:theory_eryso}

In order to apply the above model to \eryso, we only keep the zero nuclear spin isotopes (78\% of the doping ions), ignoring the $^{167}$Er isotope (22\% abundance with non-zero nuclear spin). The required smallness of the spin-spin interaction with respect to Zeeman splitting is satisfied at external magnetic field values as small as $1$~mT. In a 10 ppm \eryso crystal, at $B = 1$~mT, the Zeeman interaction varies between 5 and 50 MHz depending on the crystal orientation, whereas the Er-Er interaction ranges within 1 and 100 kHz. 
As we will discuss in \ref{discussion}, we use a very weak magnetic field to minimize the inhomongeous spin broadening. We choose to keep it constant to $B = 0.3$~mT and vary the field orientation to reveal the anisotropy.
The example of \eryso is particularly interesting because it shows strongly anisotropic properties. Indeed, in the frame $(x,y,z)$ where the $\bar{\bar{g}}$-tensor is diagonal,  
\begin{equation}
\; g_z \approx 10 \, g_y \approx 30 \, g_x \; .
\end{equation}

As pointed out above, $ \Xi\left(\bar{\bar{g}},\vec{B}\right)$ only depends on the $g_x$ and $g_y$ components, much smaller than $g_z$ in \eryso, when the applied magnetic field is directed along $Oz$. According to the general expression, derived in Appendix~\ref{sec:Vanalytical}: 
\begin{equation}
 \Xi\left(\bar{\bar{g}},\vec{B}\right) = \frac{1}{20}   \left( g_x^4 + g_y^4 - g_x^2 g_y^2 \right) \; .
 \label{eq:Vmoy_130deg}
 \end{equation}
This situation is almost reached when $\vec{B}$ is directed at about $\phi = 135^\circ$ from $D_1$ in the $D_1-D_2$ plane, or in other words, the principal axis almost lies in the $D_1-D_2$ plane .

The $g_z$ contribution raises as $\vec{B}$ direction departs from $Oz$, reaching a maximum when $\vec{B}$ lies in the $xOy$ plane. Then $\Xi\left(\bar{\bar{g}},\vec{B}\right)$ reads as: 
\begin{equation}
  \Xi\left(\bar{\bar{g}},\vec{B}\right) = 
\frac{1}{20}\left(g_z^4+g_\bot^4-g_z^2g_\bot^2\right)
  \label{eq:Vmoy_30deg}
  \end{equation}
where:
\begin{equation}
\frac{1}{g_\bot^2}=\frac{1}{B^2}\left(\frac{B_x^2}{g_y^2}+\frac{B_y^2}{g_x^2}\right) 
\; .
\end{equation}

This situation occurs for $\phi \approx 30^\circ$. According to Eqs.\ref{eq:Vmoy_130deg} and \ref{eq:Vmoy_30deg}, the relaxation rate $R$ should vary by a factor of $\left( g_z / g_y \right)^4 \simeq 10^4$ between those two situations (neglecting the small $g_x$ contribution). Note that in a scenario where the $g$-tensor is isotropic, $\Xi\left(\bar{\bar{g}},\vec{B}\right)$ would be totally independent of the external magnetic field direction.

\begin{figure}[t]
\centering
\includegraphics[width=0.9\columnwidth]{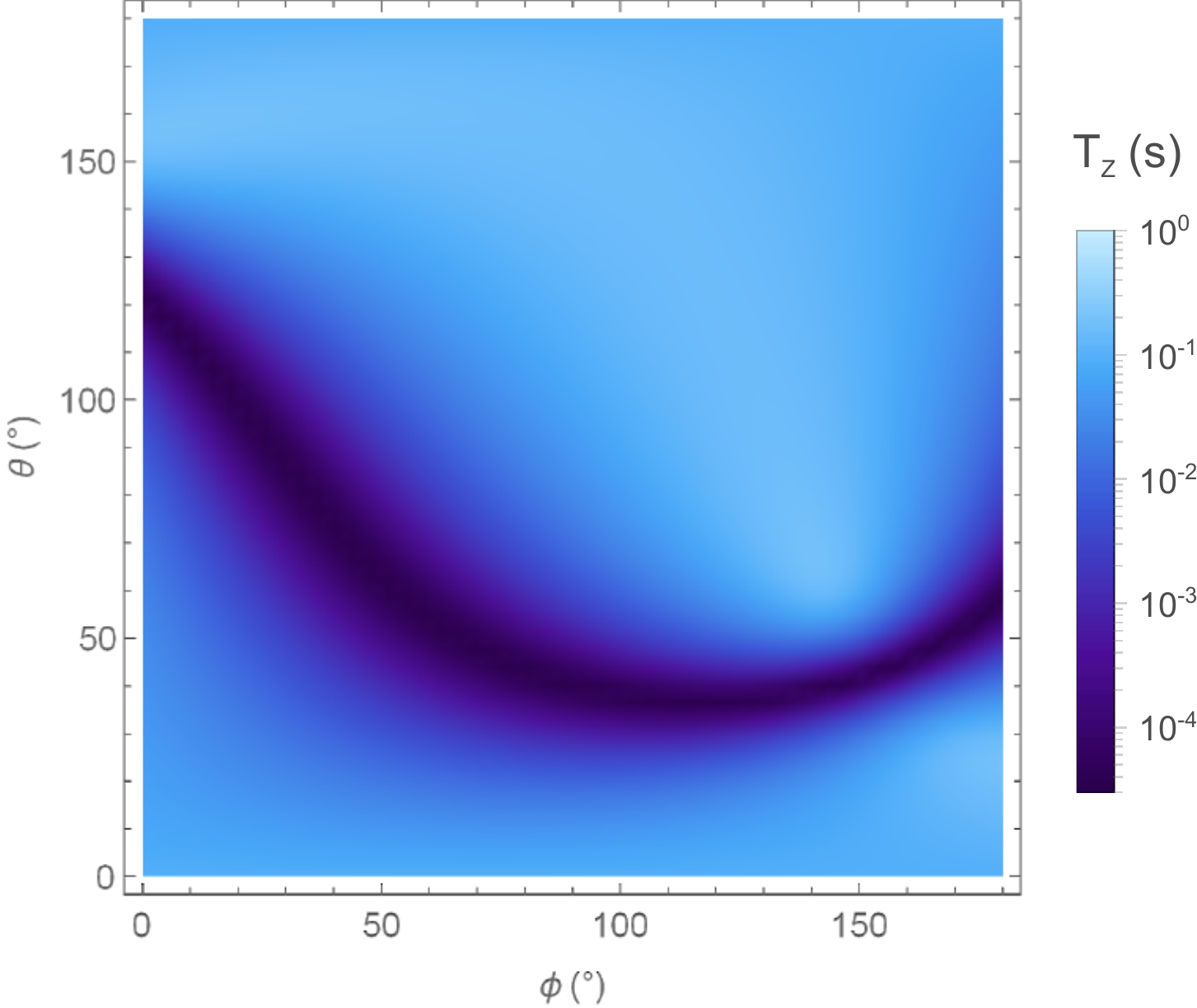} 
\caption{Map of $\displaystyle T_{ff}=1/R$ as a function of the orientation of the external magnetic field in \eryso (10 ppm). The angles $\phi$ and $\theta$ are the angles of the external magnetic field $\vec{B}$ in the crystalline frame $(D_1,D_2,b)$. When $\theta = 90^\circ$, the field lies in the $D_1-D_2$ plane. Note the logarithmic colormap.}
\label{fig:T1_map_10ppm_ErYSO}
\end{figure}

We calculate $\Xi\left(\bar{\bar{g}},\vec{B}\right)$ variations with the external magnetic field $\vec{B}$ direction in the crystalline frame $(D_1,D_2,b)$. Fig.~\ref{fig:T1_map_10ppm_ErYSO} shows the map of $\displaystyle T_{ff}=1/R$ following \eqref{eq:FGR_rate_final} as a function of $\phi$ and $\theta$ the angles of $\vec{B}$ in the $(D_1,D_2,b)$ frame. We observe strong variations within several planes, including the $D_1-D_2$ plane that we considered above.

\section{flip-flop rate in erbium doped \yso}

Because of the strong variation of the expected lifetime, different techniques must be used to cover the measurement range. We first discuss the methodology and then detail the experimental results.

\subsection{Spin lifetime optical measurement techniques}\label{techniques}

The method of choice to measure optically the spin lifetime is spectral hole burning (SHB). With modern laser stabilization techniques, one can reach SHB spectral resolution as small as the optical homogeneous linewidth, which gives access to structures below one MHz splitting. Zeeman and hyperfine structure for both Kramers and non-Kramers ions have been revealed and studied in pioneering works \cite{Macfarlane:81, PhysRevB.38.11061}. Stimulated by the recent development of diode lasers and the contextual interest in quantum information for which rare-earth crystals represent a promising quantum memory support, SHB has reappeared as a versatile tool to study the population dynamics and prepare the memory in the initial step of many storage protocols \cite{Seze_PhysRevB.73.085112, Hastings-Simon_PhysRevB.77.125111, Hastings, AFZELIUS20101566, Lauritzen_PhysRevB.85.115111, Cruzeiro}.

\begin{figure}[!ht]
\centering
\includegraphics[width=0.9\columnwidth]{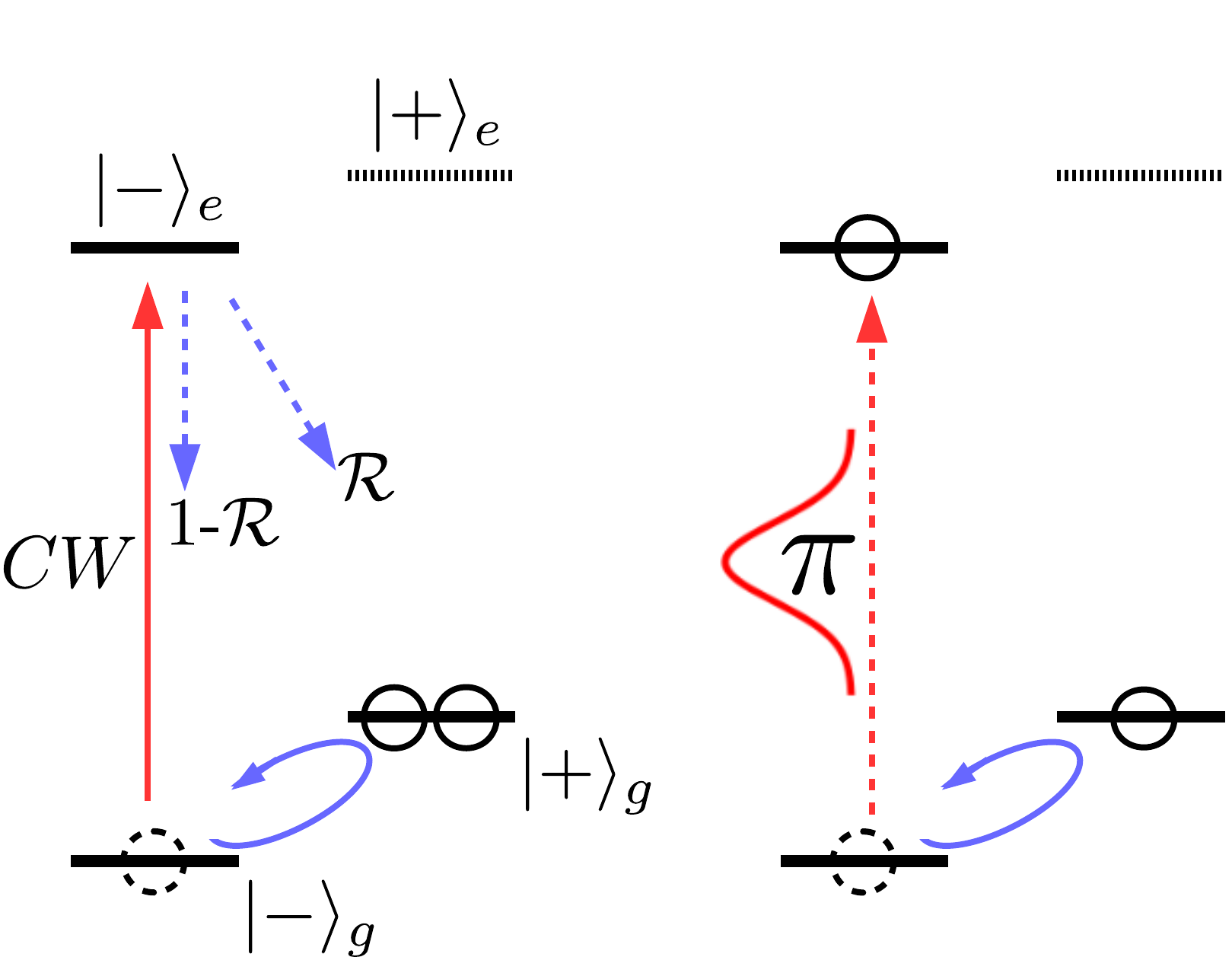}
\caption{Optical measurement techniques. Left - Accumulated spectral holeburning: the spin polarization is obtained by pumping through the optically excited state. Right - Optical inversion recovery: A spin population imbalance can also be created by transient optical excitation ($\pi$-pulse typically) when the spin lifetime is much shorter than the optical decay times. See text for details.}
\label{fig:shb_vs_pi}
\end{figure}

In a simplified picture, SHB operates as optical pumping in an equivalent three-level system, as represented in Fig.\ref{fig:shb_vs_pi} (left). This ideal scheme corresponds well to the case of \eryso. The $^{4} I_{15/2} \rightarrow \, ^{4}I_{13/2}$ optical transition of \er splits into ground and excited Kramers doublets leading to a 4-level structure. The laser frequency determines the spin levels that are addressed by the optical excitation ($\ket{-}_g \rightarrow \ket{-}_e$).  We can reduce the structure to a 3-level by just neglecting $\ket{+}_e$. This simplification is made possible because the direct spin relaxation in the excited state $\ket{-}_e \leftrightarrow \ket{+}_e$ can be neglected. More precisely, in the FF regime, the optically excited spin density is so weak that the  $\ket{-}_e \leftrightarrow \ket{+}_e$ cross-relaxation is extremely unlikely \cite{welinski2018electron, rakonjac2018spin}. The SHB structure consists of 3 levels, namely $\ket{-}_g$, $\ket{-}_e$ and $\ket{+}_g$. Instead, if the laser excites the crossed transition $\ket{-}_g \rightarrow \ket{+}_e$, which corresponds to a different subset of ions (frequency class) \cite{nilsson2004hole}, the $\ket{-}_e$ level can be neglected.

In this archetypal SHB 3-level structure, the continuous optical excitation is redistributed between the ground state sublevels, in accordance with the branching ratio $\mathcal{R}$ from the excited state. The hole decay reflects the spin relaxation. Ideally, the cumulative SHB procedure can fully polarize the spins, provided spin relaxation is slow as compared to the optical pumping rate $\mathcal{R}/T_1^\mathrm{opt}$ where $T_1^\mathrm{opt}$ is the optical lifetime. This condition may not be satisfied in Kramers ions. This is precisely the case for \er. As a consequence, the spin state preparation can be quite challenging \cite{Lauritzen_PhysRevA.78.043402}.

As we discussed in \ref{sec:theory_eryso}, the expected FF decay time varies by four orders of magnitude, and can be faster or slower than optical pumping, which is always limited by the exceptionally long value of $T_1^\mathrm{opt}=11$~ms among the rare-earth elements. When spin relaxation is too fast, one cannot accumulate spin population imbalance, and an alternative to SHB is needed. Instead of using a continuous laser for optical pumping, we perform transient optical excitation as illustrated in \figref{fig:shb_vs_pi} (right). The $\pi$-pulse achieves transient spin imbalance between $\ket{-}_g$ and $\ket{+}_g$ by promoting the $\ket{-}_g$ population into the optical excited state. The rapid recovery, caused by spin cross-relaxation, can be probed on the optical $\ket{-}_g \rightarrow \ket{-}_e$ transition. Having no time to scan the probe frequency, as we did in SHB, we instead only measure the transmission of a weak and short probe pulse. The transmission decay gives the spin lifetime. These two complementary techniques, namely accumulated SHB and optical inversion recovery ($\pi$-pulse excitation) as distinguished in \figref{fig:shb_vs_pi}, are used alternatively depending whether the expected FF lifetime is shorter or larger than $T_1^\mathrm{opt}=11$~ms. Rigorously, the relevant parameter to compare with the FF rate is the pumping rate $\mathcal{R}/T_1^\mathrm{opt}$. Nonetheless, the branching ratio  $\mathcal{R}$ also exhibits a strong magnetic field orientation dependency \cite{lauritzen}. So for simplicity, we have decided to keep $T_1^\mathrm{opt}=11$~ms as a  reference value to distinguish the different dynamical regimes.

In practice, as calculated for a 10 ppm \eryso sample (see section \ref{sec:theory_eryso}), the FF lifetime depends strongly on the magnetic field angle $\phi$ in the $D_1$-$D_2$ plane and varies between $30$~$\mu$s and $200$~ms. For $\phi \leq 70^\circ$, the FF rate is fast and so we use the optical inversion recovery measurement technique (blue shaded area in \figref{fig:Tz10ppm}). Conversely, SHB can be accumulated for $\phi > 70^\circ$ and longer spin lifetimes can be extracted from the decay curve (yellow shaded area in \figref{fig:Tz10ppm}).

\subsection{Experimental results}\label{experiment}

The experimental validation of the analytical formula \eqref{eq:FGR_rate_final} is quite challenging because of the four orders of magnitude covered by the FF rates and therefore, the spin lifetimes we need to measure. We also propose to investigate different doping concentrations to validate our formula.

The two crystals under study are monoclinic \yso, grown by Scientific Materials Corporation. Erbium substitutes for yttrium at dopant concentrations of respectively $10$ ppm and $50$ ppm. Among the two crystallographic substitution sites, we consider the one at $1536.48$ nm, referenced as site 1\cite{bottger2006spectroscopy}. We use a liquid helium cryostat to cool down the crystal at $2$ K and a superconducting coil, monitored by a low voltage supply, to generate a weak $0.3$ mT magnetic field. We use the 3 optical extinction axis ($D_1$-$D_2$-$b$) as a reference frame. Light propagates along $b$ whereas the crystal is placed on a rotating mount (Attocube ANRv51/LT) so that the external magnetic field can be rotated in the whole $D_1$-$D_2$ plane.

As discussed in \ref{techniques}, to measure the spin lifetimes of the $10$ ppm crystal, we implement the optical inversion recovery technique for angles $\phi$ from $5^\circ$ to $60^\circ$, where spin lifetimes are expected to be shorter than the optical lifetime, and a SHB experiment from $80^\circ$ to $180^\circ$ for longer lifetimes. For the $50$ ppm sample, as the rate is proportional to the density squared [see \eqref{eq:FGR_rate_final}], the lifetimes are expected to be shorter than $T_1^\mathrm{opt}=11$~ms at any angle. Therefore, the optical inversion recovery is implemented exclusively.

\begin{figure}[!t]
\centering
\includegraphics[width=0.9\columnwidth]{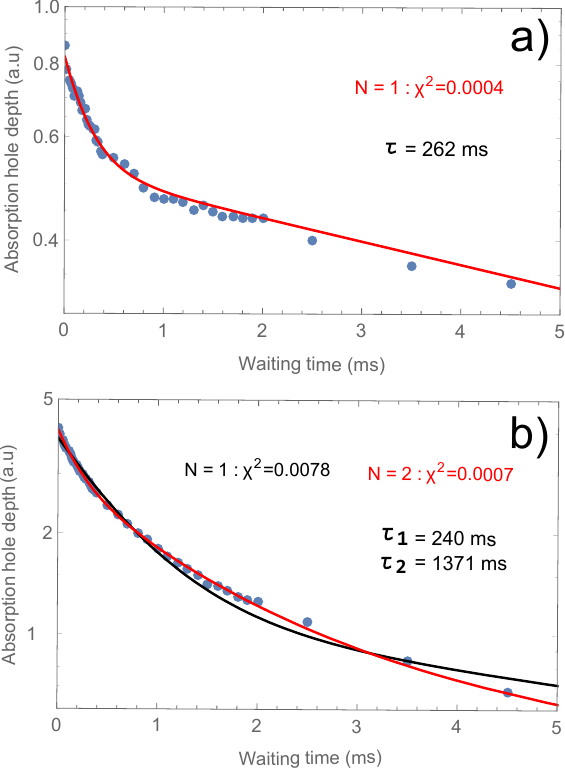}
\caption{a : Optical inversion recovery decay of the 10 ppm sample for $\vec{B}$ at $45^\circ$. We fit data with $2$ exponential decay curves ($N=1$). b : Optical inversion recovery decay of the 50 ppm sample for $\vec{B}$ at $115^\circ$. We fit data with 2 or 3 exponential decay curves (respectively black for $N=1$ and red for $N=2$).}
\label{fig:2exp10ppm}
\end{figure}

For both measurement techniques, the optical setup is the same. We split a Koheras fiber laser in two beams, the pump and the probe. The pump is amplified with a ManLight erbium-doped fiber amplifier. These two beams are temporally shaped by acousto-optic modulators, controlled by an arbitrary waveform generator (Tektronix AWG520). Then, pump and probe beams counterpropagate in the crystal with typical power of $20$ mW and $100$ $\mu$W respectively. Finally, the probe is measured by an avalanche photodetector. 
The time sequences are different for the two techniques. For SHB experiment, we use a $100$ ms pump and $500$ $\mu$s weak probe. We sweep the probe frequency to record the central hole area decays. For the optical inversion recovery, a $200$ ns rms-duration pulse has a typical area of $\pi$ (transient excitation, see \figref{fig:shb_vs_pi}). The mean transmission of a $10$ $\mu$s probe pulse gives us the temporal evolution of the absorption at different waiting times after the $\pi$-pulse (inversion recovery). \figref{fig:2exp10ppm} gives two examples of the technique for the 10 ppm amd 50 ppm samples respectively.

Both techniques exhibit decay curves characterizing the population dynamics.
In a simplified $3$-level system, the exponential decays should have two characteristic times (see \figref{fig:2exp10ppm}), $T_1^\mathrm{opt}$ and $\tau$ where $T_1^\mathrm{opt}$ is the optical lifetime and $\tau$ the spin lifetime to be compared to $T_{ff}$. We should be able to extract the spin lifetime by fitting data with the following formula 
\begin{equation}\label{eq_exp1}
\Delta_{\alpha L}(t)=\alpha L((1-a)e^{-\frac{t}{T_1^\mathrm{opt}}}+ae^{-\frac{t}{\tau}})
\end{equation}
where $T_1^\mathrm{opt}$ is fixed to $11$ ms. We choose to let the coefficients $\alpha L$ and $a$ as free parameters even thought they could be measured and calculated respectively. We actually observe that their values depend on the efficiency of the $\pi$-pulse excitation which varies significantly when rotating the sample.

\begin{figure}[!ht]
\centering
\includegraphics[width=0.9\columnwidth]{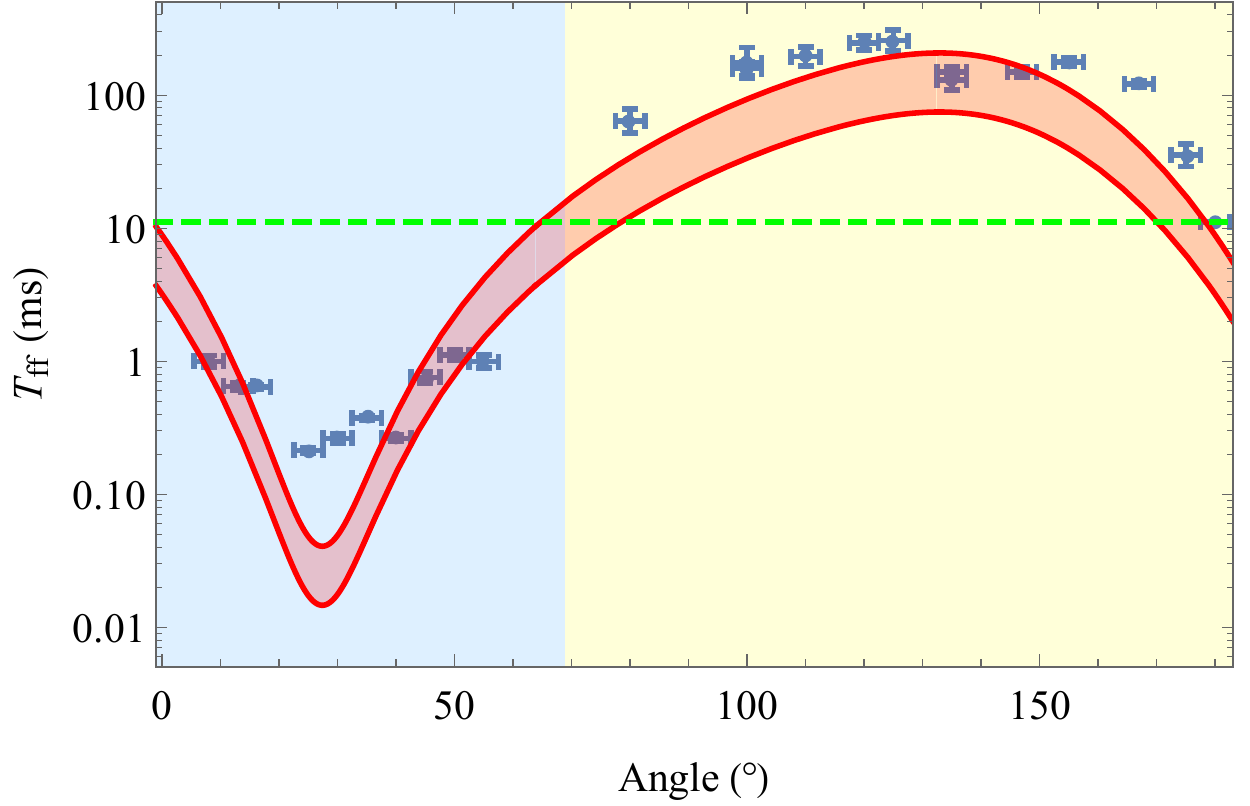}
\caption{Variation of the spin lifetime with the angle $\phi$ of the external magnetic field in the $D_1 - D_2$ plane for the $10$ ppm crystal. Blue shaded area, optical inversion recovery technique when the expected lifetime is shorter than $T_1^\mathrm{opt}=11$~ms (green dashed line). Yellow shaded area, accumulated SHB technique for longer lifetimes. The markers are the experimental measurements of $\tau_1$ and the red curves the theoretical calculation of $T_{ff}$ (see text for details)}
\label{fig:Tz10ppm}
\end{figure}

However in practice, \eqref{eq_exp1} is not sufficient to fit the very different situtations encountered when we change the magnetic field and the concentration. To obtain a good agreement between the data and the fit, we use a multi exponential decay formula as
\begin{equation}\label{eq_expN}
\Delta_{\alpha L}(t)=\alpha L\left((1-\sum\limits_{i=1}^Na_i)e^{-\frac{t}{T_1^\mathrm{opt}}}+\sum\limits_{i=1}^Na_ie^{-\frac{t}{\tau_i}}\right)
\end{equation}
and increase $N$ until $\chi^2 < 10^{-3}$ (good agreement threshold).
For the 10 ppm crystal, we reach a good agreement ($\chi^2 < 10^{-3}$) with $N=1$ for the optical inversion recovery technique (blue shaded area in \figref{fig:Tz10ppm}) as illustrated by the example in \figref{fig:2exp10ppm}). For the same sample using the SHB technique, it is necessary to increase to  $N=3$ to properly fit data, thus revealing longer timescale (typically from $1$ s to $100$ s) in this configuration as we will discuss in \ref{discussion}. For the 50 ppm sample, $N=2$ is sufficient for the different angles using optical inversion recovery as illustrated in \figref{fig:2exp10ppm}.b.

\begin{figure}[!ht]
\centering
\includegraphics[width=0.9\columnwidth]{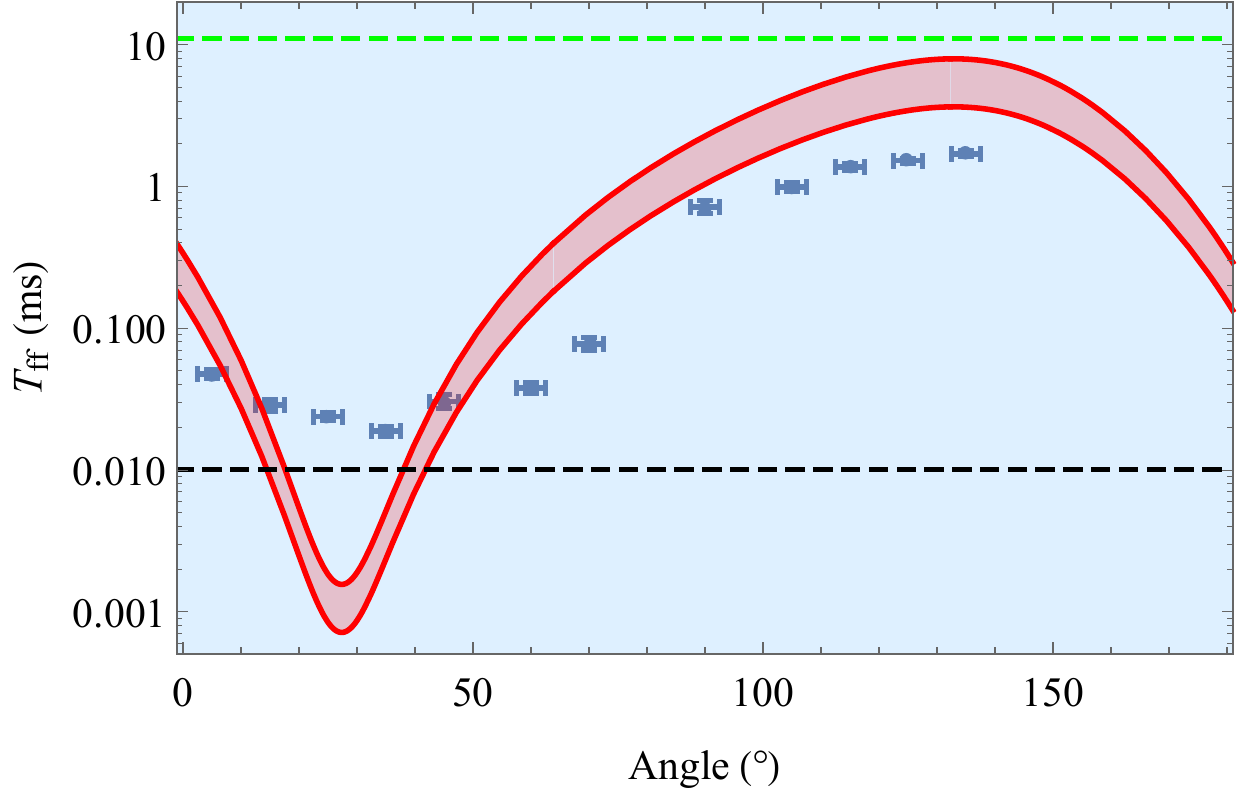}
\caption{Variation of the spin lifetime with the angle $\phi$ of the external magnetic field in the $D_1 - D_2$ plane for the $50$ ppm crystal. The experimental measurements corresponds to the optical inversion recovery technique only. We cannot measure lifetime shorter than $10$ $\mu$s (represented as a black dashed line) corresponding to our probe pulse duration.}
\label{fig:Tz50ppm}
\end{figure}

To evaluate the FF rates, we extract the characteristic time $\tau_1$ given by the dominant exponential term where $a_1$ is the largest $a_i$. The measurements for a varying magnetic field are represented in \figref{fig:Tz10ppm} and \figref{fig:Tz50ppm} for 10 ppm and 50 ppm respectively and compared with the theoretical prediction.

For the theoretical calculation, we plot a confidence interval (light-red area in Figs.\ref{fig:Tz10ppm} and \ref{fig:Tz50ppm}) corresponding to our uncertainty in the spin inhomogeneous broadening  $\Gamma_\mathrm{inh}^\mathrm{spin}$, which appears in \eqref{eq:FGR_rate_final}. $\Gamma_\mathrm{inh}^\mathrm{spin}$ varies with the magnetic field strength and orientation as will be discussed in section \ref{discussion}. We measure it for a $0.3$ mT external fields and with different orientations via SHB technique. The width of the associated antihole (positionned at the excited state splitting) corresponds to the inhomogeneous broadening of spin levels \cite{Lauritzen_PhysRevA.78.043402}. The red solid curves bounding the confidence interval have been calculated with the extremal values we observe for $\Gamma_\mathrm{inh}^\mathrm{spin}$ ($\Gamma_\mathrm{inh}^\mathrm{spin}\in \left[2.3,6.3\right]$ MHz for $10$ ppm and $\Gamma_\mathrm{inh}^\mathrm{spin}\in \left[2.8,6\right]$ MHz for $50$ ppm).

\subsection{Discussion}\label{discussion}

The agreement between the experimental data and the theoretical prediction is overall satisfying. First, we observe a strong dependence of the spin lifetime at very low magnetic field, reflecting the anisotropic character of the FF rate, as predicted by our theoretical model. From our measurements in the 10 ppm crystal, the anisotropy factor on the spin lifetime is $1.2 \times 10^3$ ($1.4 \times 10^4$ theoretically) as the magnetic field is rotated in the $D_1 - D_2$ plane.
When the lifetimes are minimum (for $\phi \approx 30^{\circ}$), the experimental values are about 10 times larger than the predictions (fact which is enhanced for the short lifetimes by the logarithmic scale). This could be explained by an uncertainty on the inhomogeneous broadening of the spin transition for instance, as discussed below. In the case of the 50 ppm crystal, the agreement is less satisfying because the ratio between the maximum and the minimum lifetimes is only $10^2$ instead of $10^4$. As we will discuss below, we cannot measure lifetimes shorter than 10 $\mu$s, still the exponential fit converges to well defined decay times. It is nevertheless difficult to track down the different decays because of the shorter orders of magnitude and the technical lower 10 $\mu$s limit. It should be also noted that this angular region ($\phi \approx 30^{\circ}$) is particularly singular because of the low $g$-factor values for ground and excited states \cite{sun2008magnetic}. In other words, the optically probed transitions $\ket{-}_g \rightarrow \ket{-}_e$ and $\ket{+}_g \rightarrow \ket{+}_e$ (see fig.\ref{fig:shb_vs_pi}) have a comparable frequency. So a mixing between different mechanisms is possible. Precautions in the analysis are needed in this angular region.
However, the overall variation caused by the change of concentration of a factor of $5^2 = 25$ is clearly observed. This validates the fact that we are in presence of a concentration dependent phenomenon, i.e. a process involving two distinct Er$^{3+}$ ions. The FF process indeed theoretically shows a quadratic dependence with the concentration. In the following we expose our experimental sources of uncertainties, as well as fundamental reasons that could explain the differences between experimental and theoretical results.



One of the main experimental sources of error is the uncertainty in measuring the spin inhomogeneous broadening $\Gamma_\mathrm{inh}^\mathrm{spin}$, which induces a significant spread of the theoretical values. The exact modeling of the spin broadening mechanism is beyond the scope of this paper but it should be noted that $\Gamma_\mathrm{inh}^\mathrm{spin}$ exhibits a specific dependency when the magnetic field is rotated. This latter can be evaluated quantitatively from the $g$-tensor \cite{Welinski2016} and would induce an extra angular dependency of the FF rate. In other words, both terms in \eqref{eq:FGR_rate_final}, $\Xi\left(\bar{\bar{g}},\vec{B}\right)$ and $\Gamma_\mathrm{inh}^\mathrm{spin}$, would depend on the magnetic field orientation. For the sake of simplicity, we plot the rate $R$ for the two measured extremal values of $\Gamma_\mathrm{inh}^\mathrm{spin}$ and keep an angular dependency of $\Xi\left(\bar{\bar{g}},\vec{B}\right)$ solely. This gives the red solid curves bounding the confidence interval in Figures~\ref{fig:Tz10ppm} and \ref{fig:Tz50ppm} (light red area). Without going much deeper into the analysis, it should be nevertheless noted than we observe at 0.3~mT approximately the same spin linewidths ($\sim 5$~MHz) than the ones measured via EPR at 60~mT \cite{Welinski2016}. Assuming a linear broadening with magnetic field, this difference corresponds to a factor of 15 in the broadening coefficients. This questions the origin of the inhomogeneous broadening, which may vary from low to high magnetic fields at different orientations. Even if we intentionally use a very weak magnetic to reduce the spin inhomogeneous broadening and measure it conveniently with the SHB technique, further investigations are needed to fully understand its origin.

Another experimental limitation imposed by our measurement techniques is the shortest measurable spin lifetimes of about 10 $\mu$s. In the optical inversion recovery technique (see Section~\ref{techniques}), we use a strong $\pi$-pulse with a duration of less than 1 $\mu$s. However, this strong pulse blinds the photo-detector for approximately 10 $\mu$s. Then, we send a 10 $\mu$s probe pulse whose integration gives the average absorption. As a consequence, the shortest spin lifetime that we can measure is 10 $\mu$s whereas the shortest expected lifetime for the 50 ppm crystal is 1 $\mu$s.

To discuss the possible influences that are not included in our model, we can first mention the presence of \er spins in the crystallographic site 2. They should not modify the FF rate of \er ions in site 1 that we probe optically. Indeed, both sites have very different spin transitions (different $g$-factors), so cross-relaxation between site 1 and 2 is off-resonant and therefore very unlikely. In other words, site 1 and 2 should behave as two independent groups. This distinction is generally true except if the sites have the same effective $g$-factor. This singular situation occurs close to $\phi \sim 135 ^\circ$ \cite{sun2008magnetic}. In that case, cross-relaxation between sites is resonant and may happen as soon as the spin transition difference (between sites) is smaller than $\Gamma_\mathrm{inh}^\mathrm{spin}$. This corresponds to a few degrees around $135 ^\circ$ and may locally increase the FF rate because the density of resonantly interacting ions is doubled. Around this region, two points of \figref{fig:Tz10ppm} seem to have a lower $T_{ff}$ value and confirm this interaction. This would deserve more investigations near $\phi \sim 135 ^\circ$. 

Another source of magnetic interaction is given by the so-called superhyperfine interaction \cite{guillot-noel_direct_2007} with the ligand (yttrium in that case) that we neglect in a first approach. The electro-nuclear spin coupling makes the electronic spin description significantly more complex, as recently discussed \cite{Car18}. Despite the weakness of the interaction between \er and the \y nuclear spin, the \er-\y proximity makes the superhyperfine interaction comparable with the \er-\er electronic coupling in diluted samples (about 100~kHz in a 10 ppm crystal). Before qualitatively incriminating any interaction that is not included in the model as the responsible of the partial disagreement with the theoretical prediction, it should be kept in mind that these interactions, including the superhyperfine coupling, are weak compared to the spin inhomogeneous broadening $\Gamma_\mathrm{inh}^\mathrm{spin}$. So they should not drastically influence the FF dynamics even if the exact coupling between \er-\y electro-nuclear mixed states still remains to be evaluated, first in a perturbative approach.

Finally, we wanted to discuss the presence of multiple characteristic decay times for the spin relaxation. As described in section \ref{experiment}, depending on the concentration and the field orientation, the population dynamics cannot be described by two timescales only, namely the optical radiative relaxation and the FF rate. We also recognize long lived structures as observed early in similar conditions by Hastings-Simon {\it at al.} \cite{Hastings}. Even if these extra mechanisms are not dominant, in the sense that they represent a small fraction of the ions, they still reveal a complex dynamics. In our SHB measurements, those long decays are visible on the anti-holes dynamics (enhanced absorption shifted by $\Delta_g - \Delta_e$ where $\Delta_g$ and $\Delta_e$ are the ground and excited state Zeeman splitting respectively). This means that the ions with the same Zeeman coefficient and sensing the same magnetic field can relax differently. This excludes for instance an explanation based on the $^{167}$Er isotope, which represents 22\% of the ions and possesses a nuclear spin. They may exhibit slower FF rates because of their much weaker effective concentration, not only because of the 22\% abundance, but also because their population is distributed among different hyperfine states making the resonant FF process much less likely. However, because of the hyperfine coupling, the shift of $\Delta_g - \Delta_e$ will be different than for the other isotopes, so $^{167}$Er do not participate to the anti-holes that we observed. Nevertheless, a comparable analysis with a 10 ppm or 50 ppm isotopically pure $^{167}$Er crystal would be interesting to explore the FF rate in the presence of hyperfine coupling.

\section{Extrapolation to other materials}

After an experimental validation in \eryso, our study can be extended to other combinations of dopants and hosts to predict the orders of magnitude of the FF rate and its angular dependence (magnetic field orientation). In this section, we compare the spin lifetime limited by FF in other materials that are investigated  in quantum information namely \ercawo \cite{bertaina_rare-earth_2007}, \erlinbo \cite{0953-4075-45-12-124013, Hastings-Simon2006} and \ndyso \cite{Cruzeiro}.

We first consider \er -doped materials to reveal the influence of the $g$-tensor anisotropy. The comparison between \ercawo and \erlinbo is particularly enlightening in that sense. For both, the \er site has a higher symmetry than in \yso.
We also analyze \ndyso to compare two Kramers ions (\er and \nd) in the anisotropic low symmetry \yso matrix. Finally we present the spin lifetime dependency as a function of the doping concentration for all these materials.

\subsection{\er ions in different host matrices}

In the previous section, we studied experimentally the FF rate in \eryso and showed its strong anisotropic behavior, as predicted by the theoretical model. \eryso is a very interesting material because of its strong anisotropic properties and long optical coherence lifetimes \cite{bottger_optical_2006}, but other host matrices are good alternatives if the FF is a limiting factor. We here focus on \ercawo and \erlinbo, which both present simpler crystalline structures and higher symmetries (S$_4$ and C$_3$ respectively for the substitution site). \ercawo exhibits long spin coherence time \cite{mims_phase_1968}. \erlinbo would benefit from all the knowledge in opto-electronics where the \linbo matrix is a reference photonic platform. As already mentioned, both materials find application in quantum information \cite{bertaina_rare-earth_2007, 0953-4075-45-12-124013, Hastings-Simon2006}.

\begin{figure}[t!]
\centering
\includegraphics[width=0.95\columnwidth]{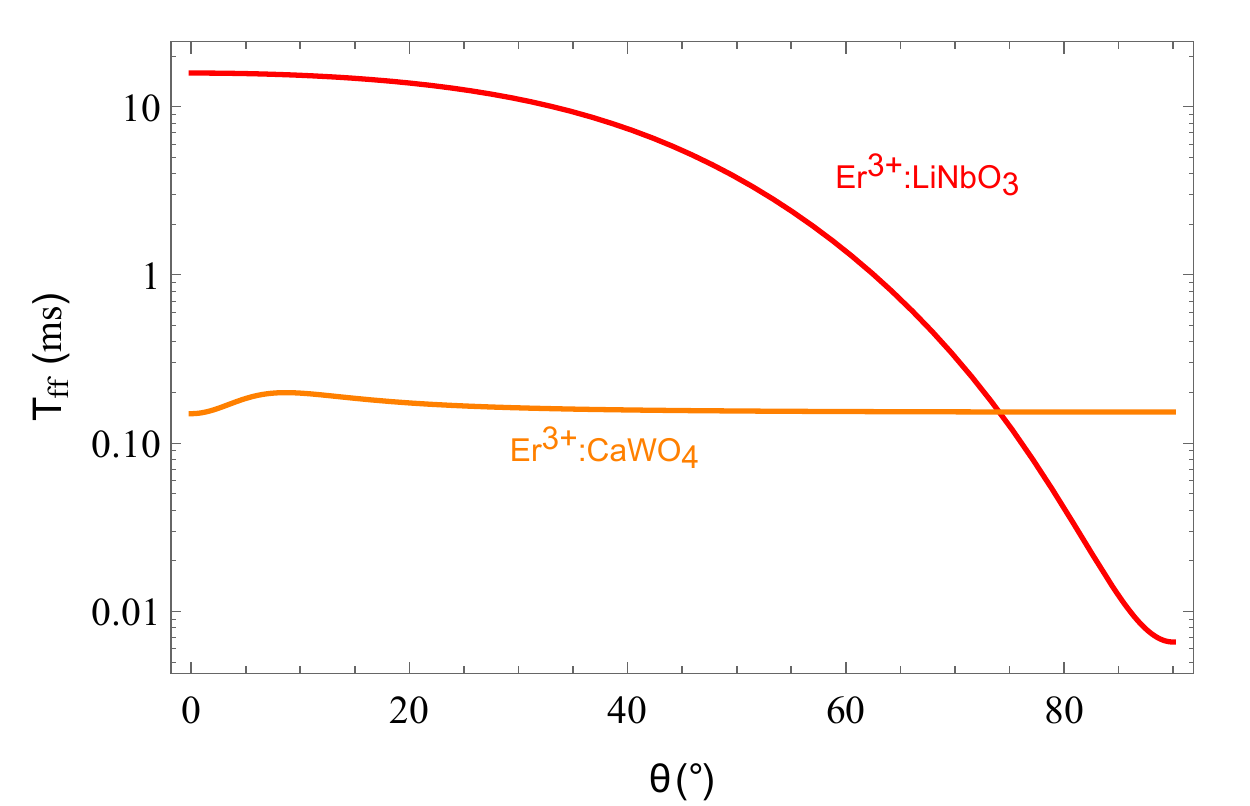}
\caption{Dependence of the spin lifetime on the orientation of the external magnetic field for \ercawo (red line) and \erlinbo (blue line) crystals. We choose a concentration of 10 ppm and a inhomogeneous spin broadening $\Gamma_{\rm inh}^{\rm spin}$ of 5 MHz in both cases.}
\label{fig:Tz_10ppm_CaWO4_LiNbO3}
\end{figure}

\ercawo and \erlinbo show a diagonal $g$-tensor in the crystalline frame $(a,b,c)$ where some properties are invariant by rotation around the $c$-axis ($a$ and $b$ are equivalent). Thus, the $g$-tensor has the following form
\begin{equation}
\bar{\bar{g}} = \left( \begin{matrix} g_\perp & 0 & 0 \\ 0 & g_\perp & 0 \\ 0 & 0 & g_\parallel \end{matrix} \right) \; .
\end{equation}
The main difference between the two materials is that $g_\perp > g_\parallel$ for \ercawo, whereas $g_\parallel > g_\perp$ for \erlinbo. This distinction has a major impact on the FF mechanism.

More precisely, the coupling term $\Xi\left(\bar{\bar{g}},\vec{B}\right)$ is almost isotropic in the case of \ercawo, i.e. almost constant when we rotate the magnetic field in the $(a,c)$ plane ($\theta$ is the angle between $\vec{B}$ and the $c$-axis). Indeed, there is always one perpendicular component that is large because $g_\perp > g_\parallel$ (see section~\ref{sec:theory_eryso}) thus the coupling $\Xi\left(\bar{\bar{g}},\vec{B}\right)$ is dominated by $g_\perp^4$ in any case.

Conversely, \erlinbo shows a very anisotropic $\Xi\left(\bar{\bar{g}},\vec{B}\right)$ because when $\vec{B} \parallel c$, the two perpendicular components are negligible. Figure~\ref{fig:Tz_10ppm_CaWO4_LiNbO3} shows the spin lifetime in these two materials as a function of the magnetic field angle $\theta$ from the $c$-axis for an \er concentration of 10 ppm. We clearly observe the quasi isotropic behavior of FF rate in \ercawo, whereas \erlinbo is strongly anisotropic. This leads to a much longer optimum spin lifetime (for $\theta = 0$) in \erlinbo than for \ercawo. We also report the maximum and minimum values of $T_{z}$ in Table~\ref{tab:comparison_mat} for a concise material comparison.

\begin{widetext}
\begin{center}
\begin{table}[t]
\centering
\begin{tabular}{>{\columncolor{Plum}}c|>{\columncolor{LightCyan}}c|>{\columncolor[gray]{0.9}}c|>{\columncolor{LightCyan}}c|>{\columncolor[gray]{0.9}}c|>{\columncolor{LightCyan}}c|>{\columncolor[gray]{0.9}}c}
 Crystal & $g_{\rm eff}^{\rm max}$  & $g_{\rm eff}^{\rm min}$ & $T_{ff}^{\rm max}$ & $T_{ff}^{\rm min}$ & $\vec{B}$ orientation for  $T_{ff}^{\rm max}$ & $\vec{B}$ orientation for  $T_{ff}^{\rm min}$   \\
\hline
\hline  
\ercawo & 8.38	& 1.25 &	199 $\mu$s & 149 $\mu$s &	$\theta = 8 ^\circ$ & $\theta = 0 ^\circ$ \\
\hline
\erlinbo & 15.14 &	2.15 &	15.8 ms &	6.55 $\mu$s &	$\theta = 0 ^\circ$ & $\theta = 90 ^\circ$ \\
\hline
\eryso & 11.7  &	1.7	& 165 ms & 	32 $\mu$s &	$\phi = 133 ^\circ$ &	$\phi = 27 ^\circ$ \\
\hline
\ndyso & 2.87 &	0.98 &	299 ms &	5.2 ms & $\phi = 104 ^\circ$ & $\phi = 12 ^\circ$ \\
\end{tabular}
\caption{Comparison between materials. For \ercawo and \erlinbo, $\theta$ is the angle of $\vec{B}$ with respect to the $c$-axis (see main text). For \eryso and \ndyso, we restrict our study to magnetic field in the $(D_1,D_2)$ plane for simplicity. The angle $\phi$ is then the angle of $\vec{B}$ in the $(D_1,D_2)$ plane (see \ref{sec:theory_eryso}). The $g$-factor value $g_{\rm eff}^{\rm max}$ (resp. $g_{\rm eff}^{\rm min}$) corresponds to the maximum (resp. minimum) $T_{ff}^{\rm max}$ lifetime (resp. $T_{ff}^{\rm min}$) }
\label{tab:comparison_mat}
\end{table}
\end{center}
\end{widetext}

\subsection{Extrapolation to \nd in \yso}

\begin{figure}[t!]
\centering
\includegraphics[width=0.95\columnwidth]{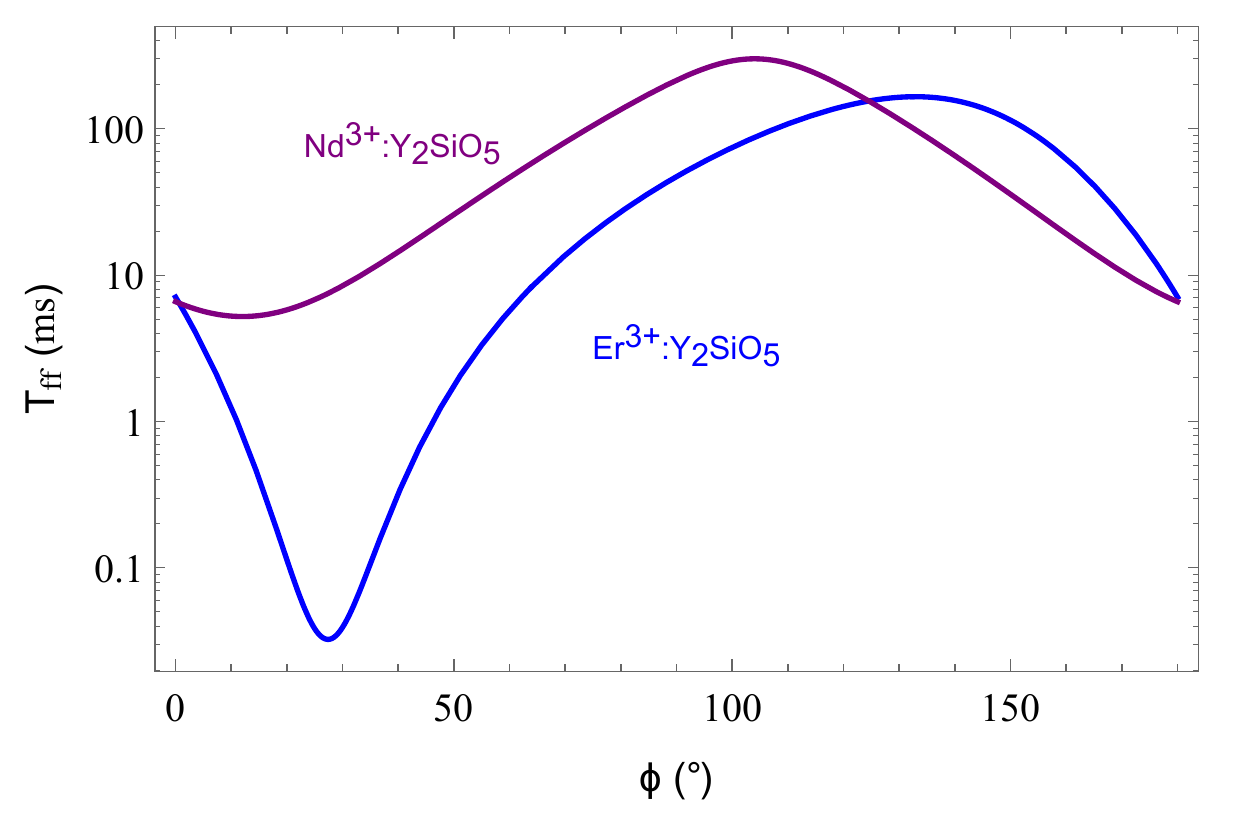}
\caption{Dependence of the spin lifetime as a function of the orientation of the external magnetic field in the $(D_1,D_2)$ plane for \ndyso (red line) and \eryso (blue line) crystals. We choose a concentration of 10 ppm and a inhomogeneous spin broadening $\Gamma_{\rm inh}^{\rm spin}$ of 5 MHz in both cases.}
\label{fig:Tz_YSO_Er_Nd}
\end{figure}

In a comparative approach, we propose to investigate the change of rare-earth ion dopant in the same matrix. We keep \yso because of its remarkable properties, (such as its strong anisotropy) and its wide use in the community. Concerning the dopant, \nd is also used in several applications. This latter is also a Kramers ions, but its $g$-tensor exhibits smaller components. Indeed, the maximum component in the principal axis is 4.17 for \ndyso \cite{wolfowicz_coherent_2015}, as compared to 14.64 for \eryso, whereas the two other components have the same order of magnitude. Figure \ref{fig:Tz_YSO_Er_Nd} shows the spin lifetime as a function of the orientation of the magnetic field in the $(D_1,D_2)$ plane for \ndyso and \eryso. This latter serves as a reference with the experimental investigation reported in Fig.\ref{fig:Tz10ppm}. We observe that the maximum is about the same, but that the minimum is larger for \ndyso. Thus, the anisotropy is enhanced for \eryso because of the very large $g$-tensor component.

\subsection{Expected concentration dependence}

\begin{figure}[t!]
\centering
\includegraphics[width=0.95\columnwidth]{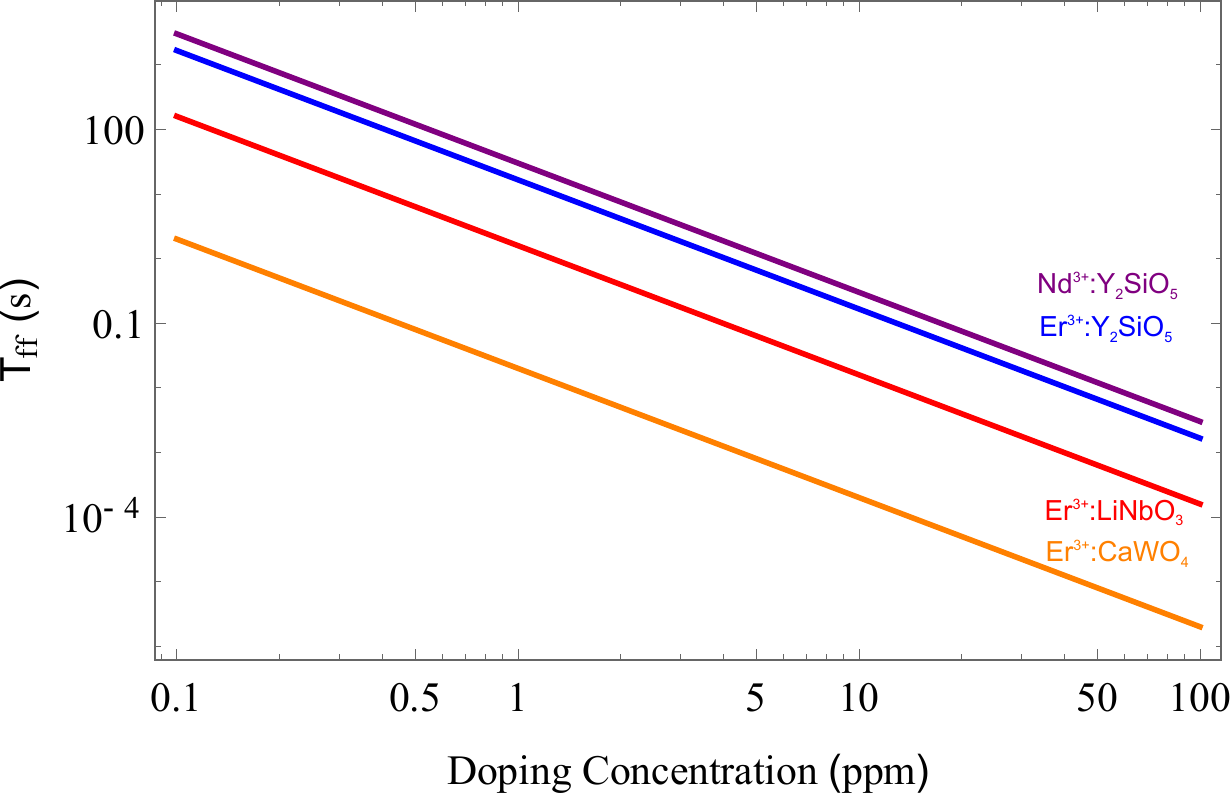}
\caption{Dependence of the spin lifetime on the doping concentration for \eryso, \ndyso, \erlinbo and \ercawo. We choose an inhomogeneous spin broadening $\Gamma_{\rm inh}^{\rm spin}$ of $5$~MHz and the best magnetic field orientation, i.e. the one which gives the longest $T_{ff}$, in all cases.}
\label{fig:Tz_vs_C}
\end{figure}

We finally present the expected concentration dependence of the spin lifetime for all the materials discussed in this work. Figure \ref{fig:Tz_vs_C} shows the lifetimes for the optimized magnetic field orientation (leading to the longest time $T_{ff}^{\rm max}$) in \eryso, \ndyso, \erlinbo and \ercawo as a function of the dopant concentration. We retrieve the orders of magnitude difference now expressed in terms of concentration. For example, if we aim at a FF-limited spin lifetime of 10~s (with the best orientation) to favor the hole-burning dynamics, this can be only achieved at concentration lower than 1.8 ppm, 1.3 ppm, 0.4 ppm and 0.05 ppm for \ndyso, \eryso, \erlinbo and \ercawo respectively. The lowest concentration level can totally prevent the optical measurements due to the lack of absorption. This aspect should be kept in mind when different combination of dopant and matrices are considered for SHB application at low magnetic field. 

\section{Conclusion}

We have shown experimentally that the erbium flip-flop dynamics is extremely anisotropic in \yso for different orientations of the magnetic field. We have verified that an order of magnitude anisotropy in the $g$-factor translates into four orders of magnitude for the flip-flop rate \cite{Cruzeiro}. In this regime, typically at low magnetic field where many experiments are operated, this scaling severly limits the spin lifetime and may directly impact the optical pumping of ensembles. Depending on the sample under consideration, a proper choice of magnetic field orientation and dopant concentration has to be made.

It should be noted that the flip-flop reduction may impose a severe additional constraint on EPR experiments when for example a large microwave coupling is targeted. For example in \eryso, a bias magnetic field minimizing the flip-flops (large $g$-factor) irremediably produces a weak coupling to the oscillating excitation field in the orthogonal direction. In other words, a proper trade-off has to be considered between cross-relaxation and an strong microwave coupling.

In any case, our study can be used as a guideline to understand the spin dymanics of other Kramers ions in different matrices, as soon as the $g$-tensor is well caracterized.

\section*{Acknowledgements}
We have received funding from the Investissements d'Avenir du LabEx PALM ExciMol and OptoRF-Er (ANR-10-LABX-0039-PALM) and from the ITMO Cancer: AVIESAN (National Alliance for Life Sciences \& Health) within the framework of the Cancer Plan.

\bibliography{FF_bib}{}

\begin{thebibliography}{49}%
\makeatletter
\providecommand \@ifxundefined [1]{%
 \@ifx{#1\undefined}
}%
\providecommand \@ifnum [1]{%
 \ifnum #1\expandafter \@firstoftwo
 \else \expandafter \@secondoftwo
 \fi
}%
\providecommand \@ifx [1]{%
 \ifx #1\expandafter \@firstoftwo
 \else \expandafter \@secondoftwo
 \fi
}%
\providecommand \natexlab [1]{#1}%
\providecommand \enquote  [1]{``#1''}%
\providecommand \bibnamefont  [1]{#1}%
\providecommand \bibfnamefont [1]{#1}%
\providecommand \citenamefont [1]{#1}%
\providecommand \href@noop [0]{\@secondoftwo}%
\providecommand \href [0]{\begingroup \@sanitize@url \@href}%
\providecommand \@href[1]{\@@startlink{#1}\@@href}%
\providecommand \@@href[1]{\endgroup#1\@@endlink}%
\providecommand \@sanitize@url [0]{\catcode `\\12\catcode `\$12\catcode
  `\&12\catcode `\#12\catcode `\^12\catcode `\_12\catcode `\%12\relax}%
\providecommand \@@startlink[1]{}%
\providecommand \@@endlink[0]{}%
\providecommand \url  [0]{\begingroup\@sanitize@url \@url }%
\providecommand \@url [1]{\endgroup\@href {#1}{\urlprefix }}%
\providecommand \urlprefix  [0]{URL }%
\providecommand \Eprint [0]{\href }%
\providecommand \doibase [0]{http://dx.doi.org/}%
\providecommand \selectlanguage [0]{\@gobble}%
\providecommand \bibinfo  [0]{\@secondoftwo}%
\providecommand \bibfield  [0]{\@secondoftwo}%
\providecommand \translation [1]{[#1]}%
\providecommand \BibitemOpen [0]{}%
\providecommand \bibitemStop [0]{}%
\providecommand \bibitemNoStop [0]{.\EOS\space}%
\providecommand \EOS [0]{\spacefactor3000\relax}%
\providecommand \BibitemShut  [1]{\csname bibitem#1\endcsname}%
\let\auto@bib@innerbib\@empty
\bibitem [{\citenamefont {Abragam}\ and\ \citenamefont
  {Bleaney}(2012)}]{abragam2012electron}%
  \BibitemOpen
  \bibfield  {author} {\bibinfo {author} {\bibfnamefont {A.}~\bibnamefont
  {Abragam}}\ and\ \bibinfo {author} {\bibfnamefont {B.}~\bibnamefont
  {Bleaney}},\ }\href@noop {} {\emph {\bibinfo {title} {Electron paramagnetic
  resonance of transition ions}}}\ (\bibinfo  {publisher} {OUP Oxford},\
  \bibinfo {year} {2012})\BibitemShut {NoStop}%
\bibitem [{\citenamefont {Grezes}\ \emph {et~al.}(2015)\citenamefont {Grezes},
  \citenamefont {Julsgaard}, \citenamefont {Kubo}, \citenamefont {Ma},
  \citenamefont {Stern}, \citenamefont {Bienfait}, \citenamefont {Nakamura},
  \citenamefont {Isoya}, \citenamefont {Onoda}, \citenamefont {Ohshima},
  \citenamefont {Jacques}, \citenamefont {Vion}, \citenamefont {Esteve},
  \citenamefont {Liu}, \citenamefont {M\o{}lmer},\ and\ \citenamefont
  {Bertet}}]{grezes}%
  \BibitemOpen
  \bibfield  {author} {\bibinfo {author} {\bibfnamefont {C.}~\bibnamefont
  {Grezes}}, \bibinfo {author} {\bibfnamefont {B.}~\bibnamefont {Julsgaard}},
  \bibinfo {author} {\bibfnamefont {Y.}~\bibnamefont {Kubo}}, \bibinfo {author}
  {\bibfnamefont {W.~L.}\ \bibnamefont {Ma}}, \bibinfo {author} {\bibfnamefont
  {M.}~\bibnamefont {Stern}}, \bibinfo {author} {\bibfnamefont
  {A.}~\bibnamefont {Bienfait}}, \bibinfo {author} {\bibfnamefont
  {K.}~\bibnamefont {Nakamura}}, \bibinfo {author} {\bibfnamefont
  {J.}~\bibnamefont {Isoya}}, \bibinfo {author} {\bibfnamefont
  {S.}~\bibnamefont {Onoda}}, \bibinfo {author} {\bibfnamefont
  {T.}~\bibnamefont {Ohshima}}, \bibinfo {author} {\bibfnamefont
  {V.}~\bibnamefont {Jacques}}, \bibinfo {author} {\bibfnamefont
  {D.}~\bibnamefont {Vion}}, \bibinfo {author} {\bibfnamefont {D.}~\bibnamefont
  {Esteve}}, \bibinfo {author} {\bibfnamefont {R.~B.}\ \bibnamefont {Liu}},
  \bibinfo {author} {\bibfnamefont {K.}~\bibnamefont {M\o{}lmer}}, \ and\
  \bibinfo {author} {\bibfnamefont {P.}~\bibnamefont {Bertet}},\ }\href
  {\doibase 10.1103/PhysRevA.92.020301} {\bibfield  {journal} {\bibinfo
  {journal} {Phys. Rev. A}\ }\textbf {\bibinfo {volume} {92}},\ \bibinfo
  {pages} {020301} (\bibinfo {year} {2015})}\BibitemShut {NoStop}%
\bibitem [{\citenamefont {Probst}\ \emph {et~al.}(2015)\citenamefont {Probst},
  \citenamefont {Rotzinger}, \citenamefont {Ustinov},\ and\ \citenamefont
  {Bushev}}]{probst}%
  \BibitemOpen
  \bibfield  {author} {\bibinfo {author} {\bibfnamefont {S.}~\bibnamefont
  {Probst}}, \bibinfo {author} {\bibfnamefont {H.}~\bibnamefont {Rotzinger}},
  \bibinfo {author} {\bibfnamefont {A.~V.}\ \bibnamefont {Ustinov}}, \ and\
  \bibinfo {author} {\bibfnamefont {P.~A.}\ \bibnamefont {Bushev}},\ }\href
  {\doibase 10.1103/PhysRevB.92.014421} {\bibfield  {journal} {\bibinfo
  {journal} {Phys. Rev. B}\ }\textbf {\bibinfo {volume} {92}},\ \bibinfo
  {pages} {014421} (\bibinfo {year} {2015})}\BibitemShut {NoStop}%
\bibitem [{\citenamefont {Williamson}\ \emph {et~al.}(2014)\citenamefont
  {Williamson}, \citenamefont {Chen},\ and\ \citenamefont
  {Longdell}}]{PhysRevLett.113.203601}%
  \BibitemOpen
  \bibfield  {author} {\bibinfo {author} {\bibfnamefont {L.~A.}\ \bibnamefont
  {Williamson}}, \bibinfo {author} {\bibfnamefont {Y.-H.}\ \bibnamefont
  {Chen}}, \ and\ \bibinfo {author} {\bibfnamefont {J.~J.}\ \bibnamefont
  {Longdell}},\ }\href {\doibase 10.1103/PhysRevLett.113.203601} {\bibfield
  {journal} {\bibinfo  {journal} {Phys. Rev. Lett.}\ }\textbf {\bibinfo
  {volume} {113}},\ \bibinfo {pages} {203601} (\bibinfo {year}
  {2014})}\BibitemShut {NoStop}%
\bibitem [{\citenamefont {O’Brien}\ \emph {et~al.}(2014)\citenamefont
  {O’Brien}, \citenamefont {Lauk}, \citenamefont {Blum}, \citenamefont
  {Morigi},\ and\ \citenamefont {Fleischhauer}}]{o2014interfacing}%
  \BibitemOpen
  \bibfield  {author} {\bibinfo {author} {\bibfnamefont {C.}~\bibnamefont
  {O’Brien}}, \bibinfo {author} {\bibfnamefont {N.}~\bibnamefont {Lauk}},
  \bibinfo {author} {\bibfnamefont {S.}~\bibnamefont {Blum}}, \bibinfo {author}
  {\bibfnamefont {G.}~\bibnamefont {Morigi}}, \ and\ \bibinfo {author}
  {\bibfnamefont {M.}~\bibnamefont {Fleischhauer}},\ }\href@noop {} {\bibfield
  {journal} {\bibinfo  {journal} {Physical review letters}\ }\textbf {\bibinfo
  {volume} {113}},\ \bibinfo {pages} {063603} (\bibinfo {year}
  {2014})}\BibitemShut {NoStop}%
\bibitem [{\citenamefont {Fernandez-Gonzalvo}\ \emph
  {et~al.}(2015)\citenamefont {Fernandez-Gonzalvo}, \citenamefont {Chen},
  \citenamefont {Yin}, \citenamefont {Rogge},\ and\ \citenamefont
  {Longdell}}]{fernandez2015coherent}%
  \BibitemOpen
  \bibfield  {author} {\bibinfo {author} {\bibfnamefont {X.}~\bibnamefont
  {Fernandez-Gonzalvo}}, \bibinfo {author} {\bibfnamefont {Y.-H.}\ \bibnamefont
  {Chen}}, \bibinfo {author} {\bibfnamefont {C.}~\bibnamefont {Yin}}, \bibinfo
  {author} {\bibfnamefont {S.}~\bibnamefont {Rogge}}, \ and\ \bibinfo {author}
  {\bibfnamefont {J.~J.}\ \bibnamefont {Longdell}},\ }\href@noop {} {\bibfield
  {journal} {\bibinfo  {journal} {Physical Review A}\ }\textbf {\bibinfo
  {volume} {92}},\ \bibinfo {pages} {062313} (\bibinfo {year}
  {2015})}\BibitemShut {NoStop}%
\bibitem [{\citenamefont {Dibos}\ \emph {et~al.}(2017)\citenamefont {Dibos},
  \citenamefont {Raha}, \citenamefont {Phenicie},\ and\ \citenamefont
  {Thompson}}]{dibos_isolating_2017}%
  \BibitemOpen
  \bibfield  {author} {\bibinfo {author} {\bibfnamefont {A.}~\bibnamefont
  {Dibos}}, \bibinfo {author} {\bibfnamefont {M.}~\bibnamefont {Raha}},
  \bibinfo {author} {\bibfnamefont {C.}~\bibnamefont {Phenicie}}, \ and\
  \bibinfo {author} {\bibfnamefont {J.}~\bibnamefont {Thompson}},\ }\href
  {http://arxiv.org/abs/1711.10368} {\bibfield  {journal} {\bibinfo  {journal}
  {arXiv:1711.10368 [physics, physics:quant-ph]}\ } (\bibinfo {year} {2017})},\
  \bibinfo {note} {arXiv: 1711.10368}\BibitemShut {NoStop}%
\bibitem [{\citenamefont {Zhong}\ \emph {et~al.}(2018)\citenamefont {Zhong},
  \citenamefont {Kindem}, \citenamefont {Bartholomew}, \citenamefont {Rochman},
  \citenamefont {Craiciu}, \citenamefont {Verma}, \citenamefont {Nam},
  \citenamefont {Marsili}, \citenamefont {Shaw}, \citenamefont {Beyer} \emph
  {et~al.}}]{zhong2018optically}%
  \BibitemOpen
  \bibfield  {author} {\bibinfo {author} {\bibfnamefont {T.}~\bibnamefont
  {Zhong}}, \bibinfo {author} {\bibfnamefont {J.~M.}\ \bibnamefont {Kindem}},
  \bibinfo {author} {\bibfnamefont {J.~G.}\ \bibnamefont {Bartholomew}},
  \bibinfo {author} {\bibfnamefont {J.}~\bibnamefont {Rochman}}, \bibinfo
  {author} {\bibfnamefont {I.}~\bibnamefont {Craiciu}}, \bibinfo {author}
  {\bibfnamefont {V.}~\bibnamefont {Verma}}, \bibinfo {author} {\bibfnamefont
  {S.~W.}\ \bibnamefont {Nam}}, \bibinfo {author} {\bibfnamefont
  {F.}~\bibnamefont {Marsili}}, \bibinfo {author} {\bibfnamefont {M.~D.}\
  \bibnamefont {Shaw}}, \bibinfo {author} {\bibfnamefont {A.~D.}\ \bibnamefont
  {Beyer}},  \emph {et~al.},\ }\href@noop {} {\bibfield  {journal} {\bibinfo
  {journal} {arXiv preprint arXiv:1803.07520}\ } (\bibinfo {year}
  {2018})}\BibitemShut {NoStop}%
\bibitem [{\citenamefont {B{\"o}ttger}\ \emph {et~al.}(2006)\citenamefont
  {B{\"o}ttger}, \citenamefont {Sun}, \citenamefont {Thiel},\ and\
  \citenamefont {Cone}}]{bottger2006spectroscopy}%
  \BibitemOpen
  \bibfield  {author} {\bibinfo {author} {\bibfnamefont {T.}~\bibnamefont
  {B{\"o}ttger}}, \bibinfo {author} {\bibfnamefont {Y.}~\bibnamefont {Sun}},
  \bibinfo {author} {\bibfnamefont {C.}~\bibnamefont {Thiel}}, \ and\ \bibinfo
  {author} {\bibfnamefont {R.}~\bibnamefont {Cone}},\ }\href@noop {} {\bibfield
   {journal} {\bibinfo  {journal} {Physical Review B}\ }\textbf {\bibinfo
  {volume} {74}},\ \bibinfo {pages} {075107} (\bibinfo {year}
  {2006})}\BibitemShut {NoStop}%
\bibitem [{\citenamefont {Sun}\ \emph {et~al.}(2008)\citenamefont {Sun},
  \citenamefont {B{\"o}ttger}, \citenamefont {Thiel},\ and\ \citenamefont
  {Cone}}]{sun2008magnetic}%
  \BibitemOpen
  \bibfield  {author} {\bibinfo {author} {\bibfnamefont {Y.}~\bibnamefont
  {Sun}}, \bibinfo {author} {\bibfnamefont {T.}~\bibnamefont {B{\"o}ttger}},
  \bibinfo {author} {\bibfnamefont {C.}~\bibnamefont {Thiel}}, \ and\ \bibinfo
  {author} {\bibfnamefont {R.}~\bibnamefont {Cone}},\ }\href@noop {} {\bibfield
   {journal} {\bibinfo  {journal} {Physical Review B}\ }\textbf {\bibinfo
  {volume} {77}},\ \bibinfo {pages} {085124} (\bibinfo {year}
  {2008})}\BibitemShut {NoStop}%
\bibitem [{\citenamefont {Milori}\ \emph {et~al.}(1995)\citenamefont {Milori},
  \citenamefont {Moraes}, \citenamefont {Hernandes}, \citenamefont {de~Souza},
  \citenamefont {Li}, \citenamefont {Terrile},\ and\ \citenamefont
  {Barberis}}]{milori_optical_1995}%
  \BibitemOpen
  \bibfield  {author} {\bibinfo {author} {\bibfnamefont {D.~M. B.~P.}\
  \bibnamefont {Milori}}, \bibinfo {author} {\bibfnamefont {I.~J.}\
  \bibnamefont {Moraes}}, \bibinfo {author} {\bibfnamefont {A.~C.}\
  \bibnamefont {Hernandes}}, \bibinfo {author} {\bibfnamefont {R.~R.}\
  \bibnamefont {de~Souza}}, \bibinfo {author} {\bibfnamefont {M.~S.}\
  \bibnamefont {Li}}, \bibinfo {author} {\bibfnamefont {M.~C.}\ \bibnamefont
  {Terrile}}, \ and\ \bibinfo {author} {\bibfnamefont {G.~E.}\ \bibnamefont
  {Barberis}},\ }\href {\doibase 10.1103/PhysRevB.51.3206} {\bibfield
  {journal} {\bibinfo  {journal} {Physical Review B}\ }\textbf {\bibinfo
  {volume} {51}},\ \bibinfo {pages} {3206} (\bibinfo {year}
  {1995})}\BibitemShut {NoStop}%
\bibitem [{\citenamefont {B\"ottger}\ \emph
  {et~al.}(2006{\natexlab{a}})\citenamefont {B\"ottger}, \citenamefont {Thiel},
  \citenamefont {Sun},\ and\ \citenamefont {Cone}}]{bottger_decoherence}%
  \BibitemOpen
  \bibfield  {author} {\bibinfo {author} {\bibfnamefont {T.}~\bibnamefont
  {B\"ottger}}, \bibinfo {author} {\bibfnamefont {C.~W.}\ \bibnamefont
  {Thiel}}, \bibinfo {author} {\bibfnamefont {Y.}~\bibnamefont {Sun}}, \ and\
  \bibinfo {author} {\bibfnamefont {R.~L.}\ \bibnamefont {Cone}},\ }\href
  {\doibase 10.1103/PhysRevB.73.075101} {\bibfield  {journal} {\bibinfo
  {journal} {Phys. Rev. B}\ }\textbf {\bibinfo {volume} {73}},\ \bibinfo
  {pages} {075101} (\bibinfo {year} {2006}{\natexlab{a}})}\BibitemShut
  {NoStop}%
\bibitem [{\citenamefont {Orbach}(1961)}]{Orbach458}%
  \BibitemOpen
  \bibfield  {author} {\bibinfo {author} {\bibfnamefont {R.}~\bibnamefont
  {Orbach}},\ }\href@noop {} {\bibfield  {journal} {\bibinfo  {journal} {Proc.
  R. Soc. Lond. A}\ }\textbf {\bibinfo {volume} {264}},\ \bibinfo {pages} {458}
  (\bibinfo {year} {1961})}\BibitemShut {NoStop}%
\bibitem [{\citenamefont {Van~Vleck}(1948)}]{van1948dipolar}%
  \BibitemOpen
  \bibfield  {author} {\bibinfo {author} {\bibfnamefont {J.}~\bibnamefont
  {Van~Vleck}},\ }\href@noop {} {\bibfield  {journal} {\bibinfo  {journal}
  {Physical Review}\ }\textbf {\bibinfo {volume} {74}},\ \bibinfo {pages}
  {1168} (\bibinfo {year} {1948})}\BibitemShut {NoStop}%
\bibitem [{\citenamefont {Portis}(1956)}]{portis1956spectral}%
  \BibitemOpen
  \bibfield  {author} {\bibinfo {author} {\bibfnamefont {A.}~\bibnamefont
  {Portis}},\ }\href@noop {} {\bibfield  {journal} {\bibinfo  {journal}
  {Physical Review}\ }\textbf {\bibinfo {volume} {104}},\ \bibinfo {pages}
  {584} (\bibinfo {year} {1956})}\BibitemShut {NoStop}%
\bibitem [{\citenamefont {Saglamyurek}\ \emph {et~al.}(2015)\citenamefont
  {Saglamyurek}, \citenamefont {Lutz}, \citenamefont {Veissier}, \citenamefont
  {Hedges}, \citenamefont {Thiel}, \citenamefont {Cone},\ and\ \citenamefont
  {Tittel}}]{saglamyurek2015efficient}%
  \BibitemOpen
  \bibfield  {author} {\bibinfo {author} {\bibfnamefont {E.}~\bibnamefont
  {Saglamyurek}}, \bibinfo {author} {\bibfnamefont {T.}~\bibnamefont {Lutz}},
  \bibinfo {author} {\bibfnamefont {L.}~\bibnamefont {Veissier}}, \bibinfo
  {author} {\bibfnamefont {M.~P.}\ \bibnamefont {Hedges}}, \bibinfo {author}
  {\bibfnamefont {C.~W.}\ \bibnamefont {Thiel}}, \bibinfo {author}
  {\bibfnamefont {R.~L.}\ \bibnamefont {Cone}}, \ and\ \bibinfo {author}
  {\bibfnamefont {W.}~\bibnamefont {Tittel}},\ }\href@noop {} {\bibfield
  {journal} {\bibinfo  {journal} {Physical Review B}\ }\textbf {\bibinfo
  {volume} {92}},\ \bibinfo {pages} {241111} (\bibinfo {year}
  {2015})}\BibitemShut {NoStop}%
\bibitem [{\citenamefont {Cruzeiro}\ \emph {et~al.}(2017)\citenamefont
  {Cruzeiro}, \citenamefont {Tiranov}, \citenamefont {Usmani}, \citenamefont
  {Laplane}, \citenamefont {Lavoie}, \citenamefont {Ferrier}, \citenamefont
  {Goldner}, \citenamefont {Gisin},\ and\ \citenamefont {Afzelius}}]{Cruzeiro}%
  \BibitemOpen
  \bibfield  {author} {\bibinfo {author} {\bibfnamefont {E.~Z.}\ \bibnamefont
  {Cruzeiro}}, \bibinfo {author} {\bibfnamefont {A.}~\bibnamefont {Tiranov}},
  \bibinfo {author} {\bibfnamefont {I.}~\bibnamefont {Usmani}}, \bibinfo
  {author} {\bibfnamefont {C.}~\bibnamefont {Laplane}}, \bibinfo {author}
  {\bibfnamefont {J.}~\bibnamefont {Lavoie}}, \bibinfo {author} {\bibfnamefont
  {A.}~\bibnamefont {Ferrier}}, \bibinfo {author} {\bibfnamefont
  {P.}~\bibnamefont {Goldner}}, \bibinfo {author} {\bibfnamefont
  {N.}~\bibnamefont {Gisin}}, \ and\ \bibinfo {author} {\bibfnamefont
  {M.}~\bibnamefont {Afzelius}},\ }\href {\doibase 10.1103/PhysRevB.95.205119}
  {\bibfield  {journal} {\bibinfo  {journal} {Phys. Rev. B}\ }\textbf {\bibinfo
  {volume} {95}},\ \bibinfo {pages} {205119} (\bibinfo {year}
  {2017})}\BibitemShut {NoStop}%
\bibitem [{\citenamefont {B{\"o}ttger}\ \emph {et~al.}(2009)\citenamefont
  {B{\"o}ttger}, \citenamefont {Thiel}, \citenamefont {Cone},\ and\
  \citenamefont {Sun}}]{bottger2009effects}%
  \BibitemOpen
  \bibfield  {author} {\bibinfo {author} {\bibfnamefont {T.}~\bibnamefont
  {B{\"o}ttger}}, \bibinfo {author} {\bibfnamefont {C.}~\bibnamefont {Thiel}},
  \bibinfo {author} {\bibfnamefont {R.}~\bibnamefont {Cone}}, \ and\ \bibinfo
  {author} {\bibfnamefont {Y.}~\bibnamefont {Sun}},\ }\href@noop {} {\bibfield
  {journal} {\bibinfo  {journal} {Physical Review B}\ }\textbf {\bibinfo
  {volume} {79}},\ \bibinfo {pages} {115104} (\bibinfo {year}
  {2009})}\BibitemShut {NoStop}%
\bibitem [{\citenamefont {Ran{\v{c}}i{\'c}}\ \emph {et~al.}(2018)\citenamefont
  {Ran{\v{c}}i{\'c}}, \citenamefont {Hedges}, \citenamefont {Ahlefeldt},\ and\
  \citenamefont {Sellars}}]{ranvcic2018coherence}%
  \BibitemOpen
  \bibfield  {author} {\bibinfo {author} {\bibfnamefont {M.}~\bibnamefont
  {Ran{\v{c}}i{\'c}}}, \bibinfo {author} {\bibfnamefont {M.~P.}\ \bibnamefont
  {Hedges}}, \bibinfo {author} {\bibfnamefont {R.~L.}\ \bibnamefont
  {Ahlefeldt}}, \ and\ \bibinfo {author} {\bibfnamefont {M.~J.}\ \bibnamefont
  {Sellars}},\ }\href@noop {} {\bibfield  {journal} {\bibinfo  {journal}
  {Nature Physics}\ }\textbf {\bibinfo {volume} {14}},\ \bibinfo {pages} {50}
  (\bibinfo {year} {2018})}\BibitemShut {NoStop}%
\bibitem [{\citenamefont {Lutz}\ \emph {et~al.}(2016)\citenamefont {Lutz},
  \citenamefont {Veissier}, \citenamefont {Thiel}, \citenamefont {Cone},
  \citenamefont {Barclay},\ and\ \citenamefont {Tittel}}]{Lutz}%
  \BibitemOpen
  \bibfield  {author} {\bibinfo {author} {\bibfnamefont {T.}~\bibnamefont
  {Lutz}}, \bibinfo {author} {\bibfnamefont {L.}~\bibnamefont {Veissier}},
  \bibinfo {author} {\bibfnamefont {C.~W.}\ \bibnamefont {Thiel}}, \bibinfo
  {author} {\bibfnamefont {R.~L.}\ \bibnamefont {Cone}}, \bibinfo {author}
  {\bibfnamefont {P.~E.}\ \bibnamefont {Barclay}}, \ and\ \bibinfo {author}
  {\bibfnamefont {W.}~\bibnamefont {Tittel}},\ }\href {\doibase
  10.1103/PhysRevA.94.013801} {\bibfield  {journal} {\bibinfo  {journal} {Phys.
  Rev. A}\ }\textbf {\bibinfo {volume} {94}},\ \bibinfo {pages} {013801}
  (\bibinfo {year} {2016})}\BibitemShut {NoStop}%
\bibitem [{\citenamefont {Yang}\ \emph
  {et~al.}(1999{\natexlab{a}})\citenamefont {Yang}, \citenamefont {Feofilov},
  \citenamefont {Williams}, \citenamefont {Milora}, \citenamefont {Tissue},
  \citenamefont {Meltzer},\ and\ \citenamefont {Dennis}}]{yang1999one}%
  \BibitemOpen
  \bibfield  {author} {\bibinfo {author} {\bibfnamefont {H.-S.}\ \bibnamefont
  {Yang}}, \bibinfo {author} {\bibfnamefont {S.}~\bibnamefont {Feofilov}},
  \bibinfo {author} {\bibfnamefont {D.~K.}\ \bibnamefont {Williams}}, \bibinfo
  {author} {\bibfnamefont {J.~C.}\ \bibnamefont {Milora}}, \bibinfo {author}
  {\bibfnamefont {B.~M.}\ \bibnamefont {Tissue}}, \bibinfo {author}
  {\bibfnamefont {R.}~\bibnamefont {Meltzer}}, \ and\ \bibinfo {author}
  {\bibfnamefont {W.}~\bibnamefont {Dennis}},\ }\href@noop {} {\bibfield
  {journal} {\bibinfo  {journal} {Physica B: Condensed Matter}\ }\textbf
  {\bibinfo {volume} {263}},\ \bibinfo {pages} {476} (\bibinfo {year}
  {1999}{\natexlab{a}})}\BibitemShut {NoStop}%
\bibitem [{\citenamefont {Yang}\ \emph
  {et~al.}(1999{\natexlab{b}})\citenamefont {Yang}, \citenamefont {Hong},
  \citenamefont {Feofilov}, \citenamefont {Tissue}, \citenamefont {Meltzer},\
  and\ \citenamefont {Dennis}}]{yang1999electron}%
  \BibitemOpen
  \bibfield  {author} {\bibinfo {author} {\bibfnamefont {H.-S.}\ \bibnamefont
  {Yang}}, \bibinfo {author} {\bibfnamefont {K.}~\bibnamefont {Hong}}, \bibinfo
  {author} {\bibfnamefont {S.}~\bibnamefont {Feofilov}}, \bibinfo {author}
  {\bibfnamefont {B.~M.}\ \bibnamefont {Tissue}}, \bibinfo {author}
  {\bibfnamefont {R.}~\bibnamefont {Meltzer}}, \ and\ \bibinfo {author}
  {\bibfnamefont {W.}~\bibnamefont {Dennis}},\ }\href@noop {} {\bibfield
  {journal} {\bibinfo  {journal} {Journal of luminescence}\ }\textbf {\bibinfo
  {volume} {83}},\ \bibinfo {pages} {139} (\bibinfo {year}
  {1999}{\natexlab{b}})}\BibitemShut {NoStop}%
\bibitem [{\citenamefont {Budoyo}\ \emph {et~al.}(2018)\citenamefont {Budoyo},
  \citenamefont {Kakuyanagi}, \citenamefont {Toida}, \citenamefont {Matsuzaki},
  \citenamefont {Munro}, \citenamefont {Yamaguchi},\ and\ \citenamefont
  {Saito}}]{Budoyo}%
  \BibitemOpen
  \bibfield  {author} {\bibinfo {author} {\bibfnamefont {R.~P.}\ \bibnamefont
  {Budoyo}}, \bibinfo {author} {\bibfnamefont {K.}~\bibnamefont {Kakuyanagi}},
  \bibinfo {author} {\bibfnamefont {H.}~\bibnamefont {Toida}}, \bibinfo
  {author} {\bibfnamefont {Y.}~\bibnamefont {Matsuzaki}}, \bibinfo {author}
  {\bibfnamefont {W.~J.}\ \bibnamefont {Munro}}, \bibinfo {author}
  {\bibfnamefont {H.}~\bibnamefont {Yamaguchi}}, \ and\ \bibinfo {author}
  {\bibfnamefont {S.}~\bibnamefont {Saito}},\ }\href
  {http://stacks.iop.org/1882-0786/11/i=4/a=043002} {\bibfield  {journal}
  {\bibinfo  {journal} {Applied Physics Express}\ }\textbf {\bibinfo {volume}
  {11}},\ \bibinfo {pages} {043002} (\bibinfo {year} {2018})}\BibitemShut
  {NoStop}%
\bibitem [{\citenamefont {Hastings-Simon}\ \emph
  {et~al.}(2008{\natexlab{a}})\citenamefont {Hastings-Simon}, \citenamefont
  {Lauritzen}, \citenamefont {Staudt}, \citenamefont {van Mechelen},
  \citenamefont {Simon}, \citenamefont {de~Riedmatten}, \citenamefont
  {Afzelius},\ and\ \citenamefont {Gisin}}]{Hastings}%
  \BibitemOpen
  \bibfield  {author} {\bibinfo {author} {\bibfnamefont {S.~R.}\ \bibnamefont
  {Hastings-Simon}}, \bibinfo {author} {\bibfnamefont {B.}~\bibnamefont
  {Lauritzen}}, \bibinfo {author} {\bibfnamefont {M.~U.}\ \bibnamefont
  {Staudt}}, \bibinfo {author} {\bibfnamefont {J.~L.~M.}\ \bibnamefont {van
  Mechelen}}, \bibinfo {author} {\bibfnamefont {C.}~\bibnamefont {Simon}},
  \bibinfo {author} {\bibfnamefont {H.}~\bibnamefont {de~Riedmatten}}, \bibinfo
  {author} {\bibfnamefont {M.}~\bibnamefont {Afzelius}}, \ and\ \bibinfo
  {author} {\bibfnamefont {N.}~\bibnamefont {Gisin}},\ }\href {\doibase
  10.1103/PhysRevB.78.085410} {\bibfield  {journal} {\bibinfo  {journal} {Phys.
  Rev. B}\ }\textbf {\bibinfo {volume} {78}},\ \bibinfo {pages} {085410}
  (\bibinfo {year} {2008}{\natexlab{a}})}\BibitemShut {NoStop}%
\bibitem [{\citenamefont {Heshami}\ \emph {et~al.}(2016)\citenamefont
  {Heshami}, \citenamefont {England}, \citenamefont {Humphreys}, \citenamefont
  {Bustard}, \citenamefont {Acosta}, \citenamefont {Nunn},\ and\ \citenamefont
  {Sussman}}]{heshami2016quantum}%
  \BibitemOpen
  \bibfield  {author} {\bibinfo {author} {\bibfnamefont {K.}~\bibnamefont
  {Heshami}}, \bibinfo {author} {\bibfnamefont {D.~G.}\ \bibnamefont
  {England}}, \bibinfo {author} {\bibfnamefont {P.~C.}\ \bibnamefont
  {Humphreys}}, \bibinfo {author} {\bibfnamefont {P.~J.}\ \bibnamefont
  {Bustard}}, \bibinfo {author} {\bibfnamefont {V.~M.}\ \bibnamefont {Acosta}},
  \bibinfo {author} {\bibfnamefont {J.}~\bibnamefont {Nunn}}, \ and\ \bibinfo
  {author} {\bibfnamefont {B.~J.}\ \bibnamefont {Sussman}},\ }\href@noop {}
  {\bibfield  {journal} {\bibinfo  {journal} {Journal of modern optics}\
  }\textbf {\bibinfo {volume} {63}},\ \bibinfo {pages} {2005} (\bibinfo {year}
  {2016})}\BibitemShut {NoStop}%
\bibitem [{\citenamefont {Asakura}\ and\ \citenamefont
  {Ando}(1998)}]{asakura1998solid}%
  \BibitemOpen
  \bibfield  {author} {\bibinfo {author} {\bibfnamefont {T.}~\bibnamefont
  {Asakura}}\ and\ \bibinfo {author} {\bibfnamefont {I.}~\bibnamefont {Ando}},\
  }\href@noop {} {\emph {\bibinfo {title} {Solid state NMR of polymers}}},\
  Vol.~\bibinfo {volume} {84}\ (\bibinfo  {publisher} {Elsevier},\ \bibinfo
  {year} {1998})\BibitemShut {NoStop}%
\bibitem [{\citenamefont {Bloembergen}(1949)}]{BLOEMBERGEN1949386}%
  \BibitemOpen
  \bibfield  {author} {\bibinfo {author} {\bibfnamefont {N.}~\bibnamefont
  {Bloembergen}},\ }\href {\doibase
  https://doi.org/10.1016/0031-8914(49)90114-7} {\bibfield  {journal} {\bibinfo
   {journal} {Physica}\ }\textbf {\bibinfo {volume} {15}},\ \bibinfo {pages}
  {386 } (\bibinfo {year} {1949})}\BibitemShut {NoStop}%
\bibitem [{\citenamefont {Dikarov}\ \emph {et~al.}(2016)\citenamefont
  {Dikarov}, \citenamefont {Zgadzai}, \citenamefont {Artzi},\ and\
  \citenamefont {Blank}}]{PhysRevApplied.6.044001}%
  \BibitemOpen
  \bibfield  {author} {\bibinfo {author} {\bibfnamefont {E.}~\bibnamefont
  {Dikarov}}, \bibinfo {author} {\bibfnamefont {O.}~\bibnamefont {Zgadzai}},
  \bibinfo {author} {\bibfnamefont {Y.}~\bibnamefont {Artzi}}, \ and\ \bibinfo
  {author} {\bibfnamefont {A.}~\bibnamefont {Blank}},\ }\href {\doibase
  10.1103/PhysRevApplied.6.044001} {\bibfield  {journal} {\bibinfo  {journal}
  {Phys. Rev. Applied}\ }\textbf {\bibinfo {volume} {6}},\ \bibinfo {pages}
  {044001} (\bibinfo {year} {2016})}\BibitemShut {NoStop}%
\bibitem [{\citenamefont {Geschwind}(1972)}]{geschwind1972electron}%
  \BibitemOpen
  \bibfield  {author} {\bibinfo {author} {\bibfnamefont {S.}~\bibnamefont
  {Geschwind}},\ }\href@noop {} {\emph {\bibinfo {title} {Electron paramagnetic
  resonance}}}\ (\bibinfo  {publisher} {Plenum Publishing Corporation},\
  \bibinfo {year} {1972})\BibitemShut {NoStop}%
\bibitem [{\citenamefont {Macfarlane}\ and\ \citenamefont
  {Shelby}(1981)}]{Macfarlane:81}%
  \BibitemOpen
  \bibfield  {author} {\bibinfo {author} {\bibfnamefont {R.~M.}\ \bibnamefont
  {Macfarlane}}\ and\ \bibinfo {author} {\bibfnamefont {R.~M.}\ \bibnamefont
  {Shelby}},\ }\href {\doibase 10.1364/OL.6.000096} {\bibfield  {journal}
  {\bibinfo  {journal} {Opt. Lett.}\ }\textbf {\bibinfo {volume} {6}},\
  \bibinfo {pages} {96} (\bibinfo {year} {1981})}\BibitemShut {NoStop}%
\bibitem [{\citenamefont {Liu}\ \emph {et~al.}(1988)\citenamefont {Liu},
  \citenamefont {Huang}, \citenamefont {Cone},\ and\ \citenamefont
  {Jacquier}}]{PhysRevB.38.11061}%
  \BibitemOpen
  \bibfield  {author} {\bibinfo {author} {\bibfnamefont {G.~K.}\ \bibnamefont
  {Liu}}, \bibinfo {author} {\bibfnamefont {J.}~\bibnamefont {Huang}}, \bibinfo
  {author} {\bibfnamefont {R.~L.}\ \bibnamefont {Cone}}, \ and\ \bibinfo
  {author} {\bibfnamefont {B.}~\bibnamefont {Jacquier}},\ }\href {\doibase
  10.1103/PhysRevB.38.11061} {\bibfield  {journal} {\bibinfo  {journal} {Phys.
  Rev. B}\ }\textbf {\bibinfo {volume} {38}},\ \bibinfo {pages} {11061}
  (\bibinfo {year} {1988})}\BibitemShut {NoStop}%
\bibitem [{\citenamefont {de~Seze}\ \emph {et~al.}(2006)\citenamefont
  {de~Seze}, \citenamefont {Louchet}, \citenamefont {Crozatier}, \citenamefont
  {Lorger\'e}, \citenamefont {Bretenaker}, \citenamefont {Le~Gou\"et},
  \citenamefont {Guillot-No\"el},\ and\ \citenamefont
  {Goldner}}]{Seze_PhysRevB.73.085112}%
  \BibitemOpen
  \bibfield  {author} {\bibinfo {author} {\bibfnamefont {F.}~\bibnamefont
  {de~Seze}}, \bibinfo {author} {\bibfnamefont {A.}~\bibnamefont {Louchet}},
  \bibinfo {author} {\bibfnamefont {V.}~\bibnamefont {Crozatier}}, \bibinfo
  {author} {\bibfnamefont {I.}~\bibnamefont {Lorger\'e}}, \bibinfo {author}
  {\bibfnamefont {F.}~\bibnamefont {Bretenaker}}, \bibinfo {author}
  {\bibfnamefont {J.-L.}\ \bibnamefont {Le~Gou\"et}}, \bibinfo {author}
  {\bibfnamefont {O.}~\bibnamefont {Guillot-No\"el}}, \ and\ \bibinfo {author}
  {\bibfnamefont {P.}~\bibnamefont {Goldner}},\ }\href {\doibase
  10.1103/PhysRevB.73.085112} {\bibfield  {journal} {\bibinfo  {journal} {Phys.
  Rev. B}\ }\textbf {\bibinfo {volume} {73}},\ \bibinfo {pages} {085112}
  (\bibinfo {year} {2006})}\BibitemShut {NoStop}%
\bibitem [{\citenamefont {Hastings-Simon}\ \emph
  {et~al.}(2008{\natexlab{b}})\citenamefont {Hastings-Simon}, \citenamefont
  {Afzelius}, \citenamefont {Min\'a\ifmmode~\check{r}\else \v{r}\fi{}},
  \citenamefont {Staudt}, \citenamefont {Lauritzen}, \citenamefont
  {de~Riedmatten}, \citenamefont {Gisin}, \citenamefont {Amari}, \citenamefont
  {Walther}, \citenamefont {Kr\"oll}, \citenamefont {Cavalli},\ and\
  \citenamefont {Bettinelli}}]{Hastings-Simon_PhysRevB.77.125111}%
  \BibitemOpen
  \bibfield  {author} {\bibinfo {author} {\bibfnamefont {S.~R.}\ \bibnamefont
  {Hastings-Simon}}, \bibinfo {author} {\bibfnamefont {M.}~\bibnamefont
  {Afzelius}}, \bibinfo {author} {\bibfnamefont {J.}~\bibnamefont
  {Min\'a\ifmmode~\check{r}\else \v{r}\fi{}}}, \bibinfo {author} {\bibfnamefont
  {M.~U.}\ \bibnamefont {Staudt}}, \bibinfo {author} {\bibfnamefont
  {B.}~\bibnamefont {Lauritzen}}, \bibinfo {author} {\bibfnamefont
  {H.}~\bibnamefont {de~Riedmatten}}, \bibinfo {author} {\bibfnamefont
  {N.}~\bibnamefont {Gisin}}, \bibinfo {author} {\bibfnamefont
  {A.}~\bibnamefont {Amari}}, \bibinfo {author} {\bibfnamefont
  {A.}~\bibnamefont {Walther}}, \bibinfo {author} {\bibfnamefont
  {S.}~\bibnamefont {Kr\"oll}}, \bibinfo {author} {\bibfnamefont
  {E.}~\bibnamefont {Cavalli}}, \ and\ \bibinfo {author} {\bibfnamefont
  {M.}~\bibnamefont {Bettinelli}},\ }\href {\doibase
  10.1103/PhysRevB.77.125111} {\bibfield  {journal} {\bibinfo  {journal} {Phys.
  Rev. B}\ }\textbf {\bibinfo {volume} {77}},\ \bibinfo {pages} {125111}
  (\bibinfo {year} {2008}{\natexlab{b}})}\BibitemShut {NoStop}%
\bibitem [{\citenamefont {Afzelius}\ \emph {et~al.}(2010)\citenamefont
  {Afzelius}, \citenamefont {Staudt}, \citenamefont {de~Riedmatten},
  \citenamefont {Gisin}, \citenamefont {Guillot-Noël}, \citenamefont
  {Goldner}, \citenamefont {Marino}, \citenamefont {Porcher}, \citenamefont
  {Cavalli},\ and\ \citenamefont {Bettinelli}}]{AFZELIUS20101566}%
  \BibitemOpen
  \bibfield  {author} {\bibinfo {author} {\bibfnamefont {M.}~\bibnamefont
  {Afzelius}}, \bibinfo {author} {\bibfnamefont {M.~U.}\ \bibnamefont
  {Staudt}}, \bibinfo {author} {\bibfnamefont {H.}~\bibnamefont
  {de~Riedmatten}}, \bibinfo {author} {\bibfnamefont {N.}~\bibnamefont
  {Gisin}}, \bibinfo {author} {\bibfnamefont {O.}~\bibnamefont
  {Guillot-Noël}}, \bibinfo {author} {\bibfnamefont {P.}~\bibnamefont
  {Goldner}}, \bibinfo {author} {\bibfnamefont {R.}~\bibnamefont {Marino}},
  \bibinfo {author} {\bibfnamefont {P.}~\bibnamefont {Porcher}}, \bibinfo
  {author} {\bibfnamefont {E.}~\bibnamefont {Cavalli}}, \ and\ \bibinfo
  {author} {\bibfnamefont {M.}~\bibnamefont {Bettinelli}},\ }\href {\doibase
  https://doi.org/10.1016/j.jlumin.2009.12.026} {\bibfield  {journal} {\bibinfo
   {journal} {Journal of Luminescence}\ }\textbf {\bibinfo {volume} {130}},\
  \bibinfo {pages} {1566 } (\bibinfo {year} {2010})},\ \bibinfo {note} {special
  Issue based on the Proceedings of the Tenth International Meeting on Hole
  Burning, Single Molecule, and Related Spectroscopies: Science and
  Applications (HBSM 2009) - Issue dedicated to Ivan Lorgere and Oliver
  Guillot-Noel}\BibitemShut {NoStop}%
\bibitem [{\citenamefont {Lauritzen}\ \emph {et~al.}(2012)\citenamefont
  {Lauritzen}, \citenamefont {Timoney}, \citenamefont {Gisin}, \citenamefont
  {Afzelius}, \citenamefont {de~Riedmatten}, \citenamefont {Sun}, \citenamefont
  {Macfarlane},\ and\ \citenamefont {Cone}}]{Lauritzen_PhysRevB.85.115111}%
  \BibitemOpen
  \bibfield  {author} {\bibinfo {author} {\bibfnamefont {B.}~\bibnamefont
  {Lauritzen}}, \bibinfo {author} {\bibfnamefont {N.}~\bibnamefont {Timoney}},
  \bibinfo {author} {\bibfnamefont {N.}~\bibnamefont {Gisin}}, \bibinfo
  {author} {\bibfnamefont {M.}~\bibnamefont {Afzelius}}, \bibinfo {author}
  {\bibfnamefont {H.}~\bibnamefont {de~Riedmatten}}, \bibinfo {author}
  {\bibfnamefont {Y.}~\bibnamefont {Sun}}, \bibinfo {author} {\bibfnamefont
  {R.~M.}\ \bibnamefont {Macfarlane}}, \ and\ \bibinfo {author} {\bibfnamefont
  {R.~L.}\ \bibnamefont {Cone}},\ }\href {\doibase 10.1103/PhysRevB.85.115111}
  {\bibfield  {journal} {\bibinfo  {journal} {Phys. Rev. B}\ }\textbf {\bibinfo
  {volume} {85}},\ \bibinfo {pages} {115111} (\bibinfo {year}
  {2012})}\BibitemShut {NoStop}%
\bibitem [{\citenamefont {Welinski}\ \emph {et~al.}(2018)\citenamefont
  {Welinski}, \citenamefont {Woodburn}, \citenamefont {Lauk}, \citenamefont
  {Cone}, \citenamefont {Simon}, \citenamefont {Goldner},\ and\ \citenamefont
  {Thiel}}]{welinski2018electron}%
  \BibitemOpen
  \bibfield  {author} {\bibinfo {author} {\bibfnamefont {S.}~\bibnamefont
  {Welinski}}, \bibinfo {author} {\bibfnamefont {P.~J.}\ \bibnamefont
  {Woodburn}}, \bibinfo {author} {\bibfnamefont {N.}~\bibnamefont {Lauk}},
  \bibinfo {author} {\bibfnamefont {R.~L.}\ \bibnamefont {Cone}}, \bibinfo
  {author} {\bibfnamefont {C.}~\bibnamefont {Simon}}, \bibinfo {author}
  {\bibfnamefont {P.}~\bibnamefont {Goldner}}, \ and\ \bibinfo {author}
  {\bibfnamefont {C.~W.}\ \bibnamefont {Thiel}},\ }\href@noop {} {\bibfield
  {journal} {\bibinfo  {journal} {arXiv preprint arXiv:1802.03354}\ } (\bibinfo
  {year} {2018})}\BibitemShut {NoStop}%
\bibitem [{\citenamefont {Rakonjac}\ \emph {et~al.}(2018)\citenamefont
  {Rakonjac}, \citenamefont {Chen}, \citenamefont {Horvath},\ and\
  \citenamefont {Longdell}}]{rakonjac2018spin}%
  \BibitemOpen
  \bibfield  {author} {\bibinfo {author} {\bibfnamefont {J.~V.}\ \bibnamefont
  {Rakonjac}}, \bibinfo {author} {\bibfnamefont {Y.-H.}\ \bibnamefont {Chen}},
  \bibinfo {author} {\bibfnamefont {S.~P.}\ \bibnamefont {Horvath}}, \ and\
  \bibinfo {author} {\bibfnamefont {J.~J.}\ \bibnamefont {Longdell}},\
  }\href@noop {} {\bibfield  {journal} {\bibinfo  {journal} {arXiv preprint
  arXiv:1802.03862}\ } (\bibinfo {year} {2018})}\BibitemShut {NoStop}%
\bibitem [{\citenamefont {Nilsson}\ \emph {et~al.}(2004)\citenamefont
  {Nilsson}, \citenamefont {Rippe}, \citenamefont {Kr{\"o}ll}, \citenamefont
  {Klieber},\ and\ \citenamefont {Suter}}]{nilsson2004hole}%
  \BibitemOpen
  \bibfield  {author} {\bibinfo {author} {\bibfnamefont {M.}~\bibnamefont
  {Nilsson}}, \bibinfo {author} {\bibfnamefont {L.}~\bibnamefont {Rippe}},
  \bibinfo {author} {\bibfnamefont {S.}~\bibnamefont {Kr{\"o}ll}}, \bibinfo
  {author} {\bibfnamefont {R.}~\bibnamefont {Klieber}}, \ and\ \bibinfo
  {author} {\bibfnamefont {D.}~\bibnamefont {Suter}},\ }\href@noop {}
  {\bibfield  {journal} {\bibinfo  {journal} {Physical Review B}\ }\textbf
  {\bibinfo {volume} {70}},\ \bibinfo {pages} {214116} (\bibinfo {year}
  {2004})}\BibitemShut {NoStop}%
\bibitem [{\citenamefont {Lauritzen}\ \emph {et~al.}(2008)\citenamefont
  {Lauritzen}, \citenamefont {Hastings-Simon}, \citenamefont {de~Riedmatten},
  \citenamefont {Afzelius},\ and\ \citenamefont
  {Gisin}}]{Lauritzen_PhysRevA.78.043402}%
  \BibitemOpen
  \bibfield  {author} {\bibinfo {author} {\bibfnamefont {B.}~\bibnamefont
  {Lauritzen}}, \bibinfo {author} {\bibfnamefont {S.~R.}\ \bibnamefont
  {Hastings-Simon}}, \bibinfo {author} {\bibfnamefont {H.}~\bibnamefont
  {de~Riedmatten}}, \bibinfo {author} {\bibfnamefont {M.}~\bibnamefont
  {Afzelius}}, \ and\ \bibinfo {author} {\bibfnamefont {N.}~\bibnamefont
  {Gisin}},\ }\href {\doibase 10.1103/PhysRevA.78.043402} {\bibfield  {journal}
  {\bibinfo  {journal} {Phys. Rev. A}\ }\textbf {\bibinfo {volume} {78}},\
  \bibinfo {pages} {043402} (\bibinfo {year} {2008})}\BibitemShut {NoStop}%
\bibitem [{\citenamefont {Lauritzen}(2010)}]{lauritzen}%
  \BibitemOpen
  \bibfield  {author} {\bibinfo {author} {\bibfnamefont {B.}~\bibnamefont
  {Lauritzen}},\ }\emph {\bibinfo {title} {Quantum memories for
  telecommunication networks}},\ \href
  {http://nbn-resolving.de/urn:nbn:ch:unige-132495} {Ph.D. thesis} (\bibinfo
  {year} {2010}),\ \bibinfo {note} {iD: unige:13249}\BibitemShut {NoStop}%
\bibitem [{\citenamefont {Welinski}\ \emph {et~al.}(2017)\citenamefont
  {Welinski}, \citenamefont {Thiel}, \citenamefont {Dajczgewand}, \citenamefont
  {Ferrier}, \citenamefont {Cone}, \citenamefont {Macfarlane}, \citenamefont
  {Chaneli\`ere}, \citenamefont {Louchet-Chauvet},\ and\ \citenamefont
  {Goldner}}]{Welinski2016}%
  \BibitemOpen
  \bibfield  {author} {\bibinfo {author} {\bibfnamefont {S.}~\bibnamefont
  {Welinski}}, \bibinfo {author} {\bibfnamefont {C.}~\bibnamefont {Thiel}},
  \bibinfo {author} {\bibfnamefont {J.}~\bibnamefont {Dajczgewand}}, \bibinfo
  {author} {\bibfnamefont {A.}~\bibnamefont {Ferrier}}, \bibinfo {author}
  {\bibfnamefont {R.}~\bibnamefont {Cone}}, \bibinfo {author} {\bibfnamefont
  {R.}~\bibnamefont {Macfarlane}}, \bibinfo {author} {\bibfnamefont
  {T.}~\bibnamefont {Chaneli\`ere}}, \bibinfo {author} {\bibfnamefont
  {A.}~\bibnamefont {Louchet-Chauvet}}, \ and\ \bibinfo {author} {\bibfnamefont
  {P.}~\bibnamefont {Goldner}},\ }\href {\doibase
  https://doi.org/10.1016/j.optmat.2016.09.039} {\bibfield  {journal} {\bibinfo
   {journal} {Optical Materials}\ }\textbf {\bibinfo {volume} {63}},\ \bibinfo
  {pages} {69 } (\bibinfo {year} {2017})},\ \bibinfo {note} {in honor of
  Professor Georges Boulon for his outstanding contributions to Optical
  Materials}\BibitemShut {NoStop}%
\bibitem [{\citenamefont {Guillot-No\"el}\ \emph {et~al.}(2007)\citenamefont
  {Guillot-No\"el}, \citenamefont {Vezin}, \citenamefont {Goldner},
  \citenamefont {Beaudoux}, \citenamefont {Vincent}, \citenamefont {Lejay},\
  and\ \citenamefont {Lorger\'e}}]{guillot-noel_direct_2007}%
  \BibitemOpen
  \bibfield  {author} {\bibinfo {author} {\bibfnamefont {O.}~\bibnamefont
  {Guillot-No\"el}}, \bibinfo {author} {\bibfnamefont {H.}~\bibnamefont
  {Vezin}}, \bibinfo {author} {\bibfnamefont {P.}~\bibnamefont {Goldner}},
  \bibinfo {author} {\bibfnamefont {F.}~\bibnamefont {Beaudoux}}, \bibinfo
  {author} {\bibfnamefont {J.}~\bibnamefont {Vincent}}, \bibinfo {author}
  {\bibfnamefont {J.}~\bibnamefont {Lejay}}, \ and\ \bibinfo {author}
  {\bibfnamefont {I.}~\bibnamefont {Lorger\'e}},\ }\href {\doibase
  10.1103/PhysRevB.76.180408} {\bibfield  {journal} {\bibinfo  {journal}
  {Physical Review B}\ }\textbf {\bibinfo {volume} {76}},\ \bibinfo {pages}
  {180408} (\bibinfo {year} {2007})}\BibitemShut {NoStop}%
\bibitem [{\citenamefont {Car}\ \emph {et~al.}(2018)\citenamefont {Car},
  \citenamefont {Veissier}, \citenamefont {Louchet-Chauvet}, \citenamefont
  {Le~Gou\"et},\ and\ \citenamefont {Chaneli\`ere}}]{Car18}%
  \BibitemOpen
  \bibfield  {author} {\bibinfo {author} {\bibfnamefont {B.}~\bibnamefont
  {Car}}, \bibinfo {author} {\bibfnamefont {L.}~\bibnamefont {Veissier}},
  \bibinfo {author} {\bibfnamefont {A.}~\bibnamefont {Louchet-Chauvet}},
  \bibinfo {author} {\bibfnamefont {J.-L.}\ \bibnamefont {Le~Gou\"et}}, \ and\
  \bibinfo {author} {\bibfnamefont {T.}~\bibnamefont {Chaneli\`ere}},\ }\href
  {\doibase 10.1103/PhysRevLett.120.197401} {\bibfield  {journal} {\bibinfo
  {journal} {Phys. Rev. Lett.}\ }\textbf {\bibinfo {volume} {120}},\ \bibinfo
  {pages} {197401} (\bibinfo {year} {2018})}\BibitemShut {NoStop}%
\bibitem [{\citenamefont {Bertaina}\ \emph {et~al.}(2007)\citenamefont
  {Bertaina}, \citenamefont {Gambarelli}, \citenamefont {Tkachuk},
  \citenamefont {Kurkin}, \citenamefont {Malkin}, \citenamefont {Stepanov},\
  and\ \citenamefont {Barbara}}]{bertaina_rare-earth_2007}%
  \BibitemOpen
  \bibfield  {author} {\bibinfo {author} {\bibfnamefont {S.}~\bibnamefont
  {Bertaina}}, \bibinfo {author} {\bibfnamefont {S.}~\bibnamefont
  {Gambarelli}}, \bibinfo {author} {\bibfnamefont {A.}~\bibnamefont {Tkachuk}},
  \bibinfo {author} {\bibfnamefont {I.~N.}\ \bibnamefont {Kurkin}}, \bibinfo
  {author} {\bibfnamefont {B.}~\bibnamefont {Malkin}}, \bibinfo {author}
  {\bibfnamefont {A.}~\bibnamefont {Stepanov}}, \ and\ \bibinfo {author}
  {\bibfnamefont {B.}~\bibnamefont {Barbara}},\ }\href {\doibase
  10.1038/nnano.2006.174} {\bibfield  {journal} {\bibinfo  {journal} {Nature
  Nanotechnology}\ }\textbf {\bibinfo {volume} {2}},\ \bibinfo {pages} {39}
  (\bibinfo {year} {2007})}\BibitemShut {NoStop}%
\bibitem [{\citenamefont {Thiel}\ \emph {et~al.}(2012)\citenamefont {Thiel},
  \citenamefont {Sun}, \citenamefont {Macfarlane}, \citenamefont {B\"ottger},\
  and\ \citenamefont {Cone}}]{0953-4075-45-12-124013}%
  \BibitemOpen
  \bibfield  {author} {\bibinfo {author} {\bibfnamefont {C.~W.}\ \bibnamefont
  {Thiel}}, \bibinfo {author} {\bibfnamefont {Y.}~\bibnamefont {Sun}}, \bibinfo
  {author} {\bibfnamefont {R.~M.}\ \bibnamefont {Macfarlane}}, \bibinfo
  {author} {\bibfnamefont {T.}~\bibnamefont {B\"ottger}}, \ and\ \bibinfo
  {author} {\bibfnamefont {R.~L.}\ \bibnamefont {Cone}},\ }\href
  {http://stacks.iop.org/0953-4075/45/i=12/a=124013} {\bibfield  {journal}
  {\bibinfo  {journal} {Journal of Physics B: Atomic, Molecular and Optical
  Physics}\ }\textbf {\bibinfo {volume} {45}},\ \bibinfo {pages} {124013}
  (\bibinfo {year} {2012})}\BibitemShut {NoStop}%
\bibitem [{\citenamefont {Hastings-Simon}\ \emph {et~al.}(2006)\citenamefont
  {Hastings-Simon}, \citenamefont {Staudt}, \citenamefont {Afzelius},
  \citenamefont {Baldi}, \citenamefont {Jaccard}, \citenamefont {Tittel},\ and\
  \citenamefont {Gisin}}]{Hastings-Simon2006}%
  \BibitemOpen
  \bibfield  {author} {\bibinfo {author} {\bibfnamefont {S.~R.}\ \bibnamefont
  {Hastings-Simon}}, \bibinfo {author} {\bibfnamefont {M.~U.}\ \bibnamefont
  {Staudt}}, \bibinfo {author} {\bibfnamefont {M.}~\bibnamefont {Afzelius}},
  \bibinfo {author} {\bibfnamefont {P.}~\bibnamefont {Baldi}}, \bibinfo
  {author} {\bibfnamefont {D.}~\bibnamefont {Jaccard}}, \bibinfo {author}
  {\bibfnamefont {W.}~\bibnamefont {Tittel}}, \ and\ \bibinfo {author}
  {\bibfnamefont {N.}~\bibnamefont {Gisin}},\ }\href {\doibase
  10.1016/j.optcom.2006.05.003} {\bibfield  {journal} {\bibinfo  {journal}
  {Optics Communications}\ }\textbf {\bibinfo {volume} {266}},\ \bibinfo
  {pages} {716} (\bibinfo {year} {2006})}\BibitemShut {NoStop}%
\bibitem [{\citenamefont {B\"ottger}\ \emph
  {et~al.}(2006{\natexlab{b}})\citenamefont {B\"ottger}, \citenamefont {Thiel},
  \citenamefont {Sun},\ and\ \citenamefont {Cone}}]{bottger_optical_2006}%
  \BibitemOpen
  \bibfield  {author} {\bibinfo {author} {\bibfnamefont {T.}~\bibnamefont
  {B\"ottger}}, \bibinfo {author} {\bibfnamefont {C.~W.}\ \bibnamefont
  {Thiel}}, \bibinfo {author} {\bibfnamefont {Y.}~\bibnamefont {Sun}}, \ and\
  \bibinfo {author} {\bibfnamefont {R.~L.}\ \bibnamefont {Cone}},\ }\href
  {\doibase 10.1103/PhysRevB.73.075101} {\bibfield  {journal} {\bibinfo
  {journal} {Physical Review B}\ }\textbf {\bibinfo {volume} {73}},\ \bibinfo
  {pages} {075101} (\bibinfo {year} {2006}{\natexlab{b}})}\BibitemShut
  {NoStop}%
\bibitem [{\citenamefont {Mims}(1968)}]{mims_phase_1968}%
  \BibitemOpen
  \bibfield  {author} {\bibinfo {author} {\bibfnamefont {W.~B.}\ \bibnamefont
  {Mims}},\ }\href {\doibase 10.1103/PhysRev.168.370} {\bibfield  {journal}
  {\bibinfo  {journal} {Physical Review}\ }\textbf {\bibinfo {volume} {168}},\
  \bibinfo {pages} {370} (\bibinfo {year} {1968})}\BibitemShut {NoStop}%
\bibitem [{\citenamefont {Wolfowicz}\ \emph {et~al.}(2015)\citenamefont
  {Wolfowicz}, \citenamefont {Maier-Flaig}, \citenamefont {Marino},
  \citenamefont {Ferrier}, \citenamefont {Vezin}, \citenamefont {Morton},\ and\
  \citenamefont {Goldner}}]{wolfowicz_coherent_2015}%
  \BibitemOpen
  \bibfield  {author} {\bibinfo {author} {\bibfnamefont {G.}~\bibnamefont
  {Wolfowicz}}, \bibinfo {author} {\bibfnamefont {H.}~\bibnamefont
  {Maier-Flaig}}, \bibinfo {author} {\bibfnamefont {R.}~\bibnamefont {Marino}},
  \bibinfo {author} {\bibfnamefont {A.}~\bibnamefont {Ferrier}}, \bibinfo
  {author} {\bibfnamefont {H.}~\bibnamefont {Vezin}}, \bibinfo {author}
  {\bibfnamefont {J.~J.}\ \bibnamefont {Morton}}, \ and\ \bibinfo {author}
  {\bibfnamefont {P.}~\bibnamefont {Goldner}},\ }\href {\doibase
  10.1103/PhysRevLett.114.170503} {\bibfield  {journal} {\bibinfo  {journal}
  {Physical Review Letters}\ }\textbf {\bibinfo {volume} {114}},\ \bibinfo
  {pages} {170503} (\bibinfo {year} {2015})}\BibitemShut {NoStop}%
\end{thebibliography}%

\appendix
\clearpage
\section{Matrix element analytical formula}
\label{sec:Vanalytical}
We choose as local frame the one where the g-tensor is diagonal, (x,y,z). The effective magnetic field can be written as
\begin{equation}
\vec{B}_{\rm eff} = \frac{1}{g_{\rm eff}} \left(
\begin{matrix} 
g_x B_x \\ g_y B_y \\ g_z B_z
\end{matrix} \right) 
= B \left(
\begin{matrix} 
\sin \Theta \cos \Phi \\ \sin \Theta \sin \Phi \\ \cos \Theta
\end{matrix} \right)  \; ,
\label{eq:Beff_vector}
\end{equation}
where $B = ||\vec{B}||$, $g_{\rm eff} =  \sqrt{g_x^2 B_x^2 + g_y^2 B_y^2 + g_z^2 B_z^2} /B $ and $(\Phi,\Theta)$ are the angular coordinates of $\vec{B}_{\rm eff}$ in the $(x,y,z)$ frame. Then, we can write the the $2 \times 2$ Zeeman Hamiltonian as
\begin{equation}
H_{{\rm Z}} = - \mu_{\rm B}g_{\rm eff}\:B \left( \sin \Theta \cos \Phi S_x +\sin \Theta \sin \Phi S_y +\cos \Theta S_z \right)  \; .
\end{equation}
The eigenvalues of $H_{{\rm Z}}$ are $\pm \frac{1}{2}\mu_{\rm B}  g_{\rm eff} B$ and the eigenstates are
\begin{align}
\begin{split}
\ket{+} &= \cos \frac{\Theta}{2} \ket{+1/2} + \sin \frac{\Theta}{2} e^{i \Phi} \ket{-1/2} \; , \\
\ket{-} &= \cos \frac{\Theta}{2} \ket{-1/2} - \sin \frac{\Theta}{2} e^{-i \Phi} \ket{+1/2} \; .
\label{eq:plus_minus}
\end{split}
\end{align}
We aim at calculating the dimension-less factor $\Xi\left(\bar{\bar{g}},\vec{B}\right)$ (see Eq.\ref{eq:Xi_factor}). Expressing the unit vector $\vec{u}_{ij}$ in terms of angular coordinates as: 
\begin{equation}
\vec{u}_{ij} = \frac{\vec{r}_{ij}}{r_{ij}}  = \left(
\begin{matrix} 
\sin \theta \cos \phi \\ \sin \theta \sin \phi \\ \cos \theta
\end{matrix} \right)  \; ,
\label{eq:rij_vector}
\end{equation}
we readily obtain 
\begin{widetext}
\begin{align}
\begin{split}
\mathcal{A}(\phi, \theta) & = 3 \left\langle -+ | \left( \vec{\mu}_i \cdot \vec{r}_{ij} \right) \left( \vec{\mu}_j \cdot \vec{r}_{ij} \right) | +- \right\rangle /\mu_{\rm B}^2\\
& = \frac{3}{4}  \left[ \sin^2 \theta \cos^2 \Theta \, \mathcal{R}^2 + g_z^2 \cos^2 \theta \sin^2 \Theta - 2 g_z \sin \theta \cos \theta \sin \Theta \cos \Theta \, \mathcal{R} + \sin^2 \theta \, \mathcal{I}^2 \right] \; ,
\label{eq:mui_dot_r_element}
\end{split}
\end{align}
where $\Gamma= \left( g_x \cos \phi + i g_y \sin \phi \right) e^{-i \Phi} = \mathcal{R} + i \mathcal{I}$, and:
\begin{equation}
\mathcal{B} = \left\langle -+ | \vec{\mu}_i \cdot \vec{\mu}_j | +- \right\rangle /\mu_{\rm B}^2= \frac{1}{8}  \left[ 2 \left( g_x^2 + g_y^2 \right) - \sin^2 \Theta \left( g_x^2 + g_y^2 - 2 g_z^2 \right) - \sin^2 \Theta \cos(2\Phi) \left( g_x^2 - g_y^2 \right) \right] \; .
\label{eq:mui_dot_muj_element}
\end{equation}
\end{widetext}
Since $\mathcal{A}(\phi, \theta)$ and $\mathcal{B}$ are real, factor $\Xi\left(\bar{\bar{g}},\vec{B}\right)$ can be expressed as:
\begin{equation}
\Xi\left(\bar{\bar{g}},\vec{B}\right)=\left\langle \left[\left( \mathcal{A}(\phi, \theta) - \mathcal{B} \right)\right]^2\right\rangle_{\theta,\phi},
\end{equation}
where 
\begin{equation}
\left\langle f\left(\theta,\phi\right)\right\rangle_{\theta,\phi}=\frac{1}{4\pi}\int_0^{2\pi}d\phi\int_0^{\pi}\sin\theta d\theta f\left(\theta,\phi\right)
\end{equation}
Noticing that $\left\langle \mathcal{A}(\phi, \theta) \right\rangle _ {\phi, \theta} = \mathcal{B}$, we expand $\Xi\left(\bar{\bar{g}},\vec{B}\right)$ as:
\begin{align}
\begin{split}
\Xi\left(\bar{\bar{g}},\vec{B}\right) & = \left\langle \left[ \mathcal{A}(\phi, \theta) \right]^2 \right\rangle _ {\phi, \theta} + \mathcal{B} \left( \mathcal{B} -2 \left\langle \mathcal{A}(\phi, \theta) \right\rangle _ {\phi, \theta} \right) \\ & =\left\langle \left[ \mathcal{A}(\phi, \theta) \right]^2 \right\rangle _ {\phi, \theta} - \mathcal{B}^2 \; .
\label{eq:V2_averaged}
\end{split}
\end{align}
According to Eq.~\ref{eq:mui_dot_muj_element}, $\mathcal{B}$ can be expressed in terms of the three parameters 
\begin{align}
& S  =  g_x^2+g_y^2 \; , \nonumber \\
& \Delta_1  =  g_x^2+g_y^2 - 2g_z^2 \; , \label{eq:anisotropy_parameters} \\
& \Delta_2  =  g_x^2-g_y^2 \; . \nonumber
\end{align} 
that characterize the anisotropy of the $g$-tensor. The anisotropy between the longitudinal and transverse axis, and within the two transverse axis, is respectively represented by $\Delta_1$ and $\Delta_2$. In the case of an isotropic $g$-tensor, $S= 2 g_{\rm eff}^2$, and $\Delta_1 = \Delta_2 = 0$. 

In the same way $\left\langle \left[ \mathcal{A}(\phi, \theta) \right]^2 \right\rangle _ {\phi, \theta}$ can be expressed in terms of those anisotropy parameters: 
\begin{widetext}
\begin{align}
\begin{split}
\left\langle \left[ \mathcal{A}(\phi, \theta) \right]^2 \right\rangle _ {\phi, \theta} = & \frac{9}{80} \left\lbrace \left[ \frac{2}{3} \, S - \frac{1}{2} \sin^2 \Theta \, \Delta_1 + \frac{1}{2} \left( \cos^2 \Theta - \frac{1}{3} \right) \cos \left( 2 \Phi \right) \Delta_2 \right]^2 \right. \\ 
& \left.
+ \frac{2}{9} \left( S - \cos \left( 2 \Phi \right) \Delta_2 \right)^2 + \frac{1}{3} \cos^2 \Theta \, \sin^2 \left( 2 \Phi \right) \Delta_2^2 \right \rbrace  \; .
\label{eq:A2_averaged}
\end{split}
\end{align}

\end{widetext}

\section{Weighted enumeration of the interacting spins}
\label{sec:sum_distance}
We here enumerate the number of interacting spins. The interaction strength decreases like 
${r_{k}^6}$ where $r_{k}$ is the distance between two spins but the number of interacting pairs increases. A proper enumeration of $\sum_{k} \frac{1}{r_{k}^6}$ is necessary.

We consider a fictitious cubic arrangement of spins of density $n_s $. The cube side is $a=\sqrt[3]{n_s}$ and each spin is located at a frame node referenced by its coordinated $(ma,na,pa)$ where $m,n,p$ are signed integers.
\begin{enumerate}
\item The first layer corresponds to $m=\pm1$ or $n=\pm1$ or $p=\pm1$ i.e. $m^2+n^2+p^2=1$ (6 possibilities with distance $a$) whose contribution to $\sum_{k} \frac{1}{r_{k}^6}$ is $6 \, n_s^2$.
\item The second layer corresponds to $m^2+n^2+p^2=2$ (12 possibilities with distance $\sqrt{2}a$) whose contribution is $\frac{3}{2} \, n_s^2$.
\item The third layer $m^2+n^2+p^2=3$ with 8 possibilities at distance $\sqrt{3}a$ contributes as $\frac{8}{27} \, n_s^2$.
\end{enumerate}
The recursive enumeration is pushed to the 10$^\mathrm{th}$ layer where the sum converges to
$\sum_{k} \frac{1}{r_{k}^6} = 8.4 \, n_s^2$





\end{document}